\documentclass[%
  reprint,
 amsmath,
 amssymb,
 aps,
 prstab,
]{revtex4-1}

\usepackage{graphicx}

\begin{document}


\title{Explicit Symplectic Integrator for Particle Tracking in $s$-dependent Static Electric and Magnetic Fields with Curved Reference Trajectory}

\thanks{Work supported by the Science and Technology Facilities Council, UK.}

\author{A.\,Wolski}
\email{a.wolski@liverpool.ac.uk}
\author{A.\,Herrod}
\affiliation{University of Liverpool, Liverpool, and the Cockcroft Institute, Warrington, UK.}

\date{\today}

\begin{abstract}
We describe a method for symplectic tracking of charged particles through static electric and magnetic
fields.  The method can be applied to cases where the fields have a dependence on longitudinal as well
as transverse position, and where the reference trajectory may have non-zero curvature.  Application
of the method requires analytical expressions for the scalar and vector potentials: we show how suitable
expressions, in the form of series analogous to multipole expansions, can be constructed from numerical
field data, allowing the method to be used in cases where only numerical field data are available.
\end{abstract}

\maketitle





\section{Introduction}
The magnetic and (in some cases) electric fields used to guide particles in an accelerator are often arranged
so that particles ideally follow a curved trajectory.  In simple cases, for example a magnetic dipole field, standard
expressions can be used to calculate the path of a particle through both the main field and the fringe field regions
of the relevant element.  However, in more complex cases, calculating particle trajectories can be challenging:
such cases include, for example, situations where quadrupole or higher-order multipole fields are included by
design within a dipole field, or where account needs to be taken of multipole components occuring from systematic
or random errors within the element.  In general, the problem of particle tracking can be broken down into two
parts.  First, an accurate description of the field is needed; and second, the equations of motion through the
field must be integrated to find the path followed by a given particle.  It is often possible to use a numerical
field map to describe the field; then, standard integration algorithms (for example, Runge--Kutta algorithms)
can be used to integrate the equations of motion.
However, an approach such as this can be computationally expensive, both in terms of the memory needed
to store the field data, and in terms of the processing involved in integrating the equations of motion.  Furthermore,
if there are specific constraints or requirements for the trajectories, then additional challenges can occur.  For example,
if the tracking must obey the symplectic condition, then an explicit Runge--Kutta integration algorithm cannot be used.
Symplectic Runge--Kutta algorithms do exist, but are implicit in the sense that each step requires the solution of a
set of algebraic equations that can add significantly to the computation time.

Regarding the description of the field, an alternative approach to a numerical field map is to represent the field
as a superposition of a number of ``modes''.  Given a set of coefficients, the field can be calculated at any
position by summing the functions describing the different modes.  This is the approach generally taken for
multipole fields, for example, where the horizontal and vertical magnetic field components $B_x$ and $B_y$ (respectively)
are given by:
\begin{equation}
B_y + i B_x = \sum_{m = 0}^{m_\mathrm{max}} C_m (x + i y)^m.
\label{multipoleexpansion}
\end{equation}
The upper limit of the sum, $m_\mathrm{max}$ is chosen to provide the accuracy required for the field.  The
advantages of this approach over a numerical field map are first, that the data describing the field are contained
in a relatively small set of coefficients, and second, that the calculation of the field at an arbitrary point does not
need interpolation between grid points, which can be an issue in some circumstances for a numerical field map.
The field represented by the multipole expansion (\ref{multipoleexpansion}) is independent of the distance along
the reference trajectory, and so is appropriate for the main field region within an accelerator element.  Depending
on the situation being considered, fringe fields
may be neglected altogether (as in the ``hard edge'' approximation), or may be represented using appropriate
expressions based, for example, on generalised gradients \cite{dragt} or formulae representing
solutions to Maxwell's equations with appropriate limiting behaviour \cite{muratoriwolskijones}.

A semi-analytical field description such as (\ref{multipoleexpansion}) has a further advantage over a purely
numerical description in the context of particle tracking.  In some cases, it is possible to construct explicit transfer
maps parameterised, for example, in terms of the mode coefficients and element length: the transfer maps then
offer the possibility of greater computational efficiency over numerical integration techniques, such as Runge--Kutta
algorithms.  Furthermore, if the transfer maps are constructed in an appropriate way, then the tracking can satisfy
requirements such as symplecticity.  An explicit symplectic integrator for general $s$-dependent static magnetic
fields, in systems with a straight reference trajectory, has been presented by Wu, Forest and Robin \cite{wfrintegrator}.
Application of the integrator requires the derivatives of the vector potential; it is therefore convenient to have
a semi-analytical field description, which allows the derivatives to be expressed in terms of appropriate modes in
the same way as the potential itself, thus avoiding the need for taking derivatives numerically.

In elements designed to bend the beam trajectory, it is usually convenient to use a reference trajectory that
follows the intended curvature of the path followed by the beam.  In such cases, the standard multipole expansion
(\ref{multipoleexpansion}) must be modified to give a field that satisfies Maxwell's equations.  For completeness,
we would like to have a set of modes that can be used to describe three-dimensional electric and magnetic fields
in a co-ordinate system based on a curved reference trajectory, and an efficient method for integrating the equations of
motion for particles moving through these fields.  In this paper, we present a suitable set of modes for static
electric and magnetic fields,
and an explicit symplectic integrator for tracking particles through a given field (i.e.~a field represented by a certain
set of coefficients).  The mode decomposition that we use is based on solutions to Laplace's equation in toroidal
co-ordinates; the explicit symplectic integrator is developed following the method of Wu, Forest and Robin
\cite{wfrintegrator}.

\section{Definitions}
We consider a particle of charge $q$ moving (at a relativistic velocity $v$) through a static electromagnetic
field described by a scalar potential $\Phi$ and a vector potential $\mathbf{A} = (A_x, A_y, A_s)$.  The Hamiltonian
for the motion of the particle is \cite{wolski}:
\begin{widetext}
\begin{equation}
H = \frac{\delta}{\beta_0} - (1 + h x)\sqrt{ \left( \delta + \frac{1}{\beta_0} - \frac{q\Phi}{c P_0} \right)^2
- (p_x - a_x)^2 - (p_y - a_y)^2 - \frac{1}{\beta_0^2 \gamma_0^2} } - (1 + h x)a_s,
\label{hamiltonian}
\end{equation}
\end{widetext}
where a particle with the chosen reference momentum $P_0$ has velocity $\beta_0 c$ and
relativistic factor $\gamma_0 = (1 - \beta_0)^{-\frac{1}{2}}$, and the scaled vector potential
$\mathbf{a} = (a_x, a_y, a_s) = q\mathbf{A} / P_0$.  The independent variable for the
system is $s$, corresponding to distance along a reference trajectory.  The reference trajectory follows
the arc of a circle (in the plane perpendicular to $y$) with radius $\rho = 1/h$.  At any point along
the reference trajectory, the co-ordinates $x$ and $y$ describe (respectively) the horizontal and vertical position
of the particle in a plane perpendicular to the reference trajectory.  The longitudinal co-ordinate is defined:
\begin{equation}
z = \frac{s}{\beta_0} - ct,
\end{equation}
where the particle arrives at position $s$ along the reference trajectory at time $t$ (and we can
assume that for the reference particle, $s = 0$ at time $t = 0$).

The momenta conjugate to the co-ordinates $x$ and $y$ are:
\begin{equation}
p_x = \frac{\gamma m v_x}{P_0} + a_x, \qquad
p_y = \frac{\gamma m v_y}{P_0} + a_y,
\end{equation}
where $\gamma$ is the relativistic factor of the particle, $m$ is the mass, and
$v_x$ and $v_y$ are the components of the velocity parallel to the $x$ and $y$ axes.
The longitudinal conjugate momentum is:
\begin{equation}
\delta = \frac{E}{c P_0} - \frac{1}{\beta_0},
\end{equation}
where $E = \gamma m c^2 + q \Phi$ is the total energy of the particle.
To simplify some of the formulae, we introduce the ``scaled'' scalar potential $\phi$, defined
by:
\begin{equation}
\phi = \frac{q \Phi}{c P_0}.
\end{equation}
\vspace*{0.2cm}

\section{Derivation of the symplectic integrator\label{sectionderivation}}
Our method follows the technique of Wu, Forest and Robin \cite{wfrintegrator}.
We first extend phase space by introducing a new independent variable $\sigma$, so
that $s$ is now a dynamical variable with conjugate momentum $p_s$.  The Hamiltonian describing
the motion of a particle through an electrostatic field with scaled potential $\phi = \phi(x,y,s)$
and magnetic field described by a scaled potential $\mathbf{a}=(a_x,a_y,a_s)$ is now:
\begin{widetext}
\begin{equation}
H^\prime = p_s + \frac{\delta}{\beta_0} - (1 + h x)\sqrt{ \left( \delta + \frac{1}{\beta_0} - \phi \right)^2
- (p_x - a_x)^2 - (p_y - a_y)^2 - \frac{1}{\beta_0^2 \gamma_0^2} } - (1 + hx)a_s.
\label{hamiltonianextendedps}
\end{equation}
\end{widetext}
We shall consider the special case where the magnetic field has a uniform vertical field component, which
can be represented by a component of the vector potential:
\begin{equation}
a_s = -k_0 x + \frac{k_0 h x^2}{2(1+h x)},
\label{magneticvectorpotential}
\end{equation}
where $k_0 = qB_0/P_0$ for a magnetic field of strength $B_0$.  If the field is correctly matched to the
curvature of the reference trajectory (so that the reference trajectory is a possible physical trajectory of
a particle with momentum $P_0$), then $h = k_0$.  Other components of the magnetic field can be
included in the components $a_x$ and $a_y$ of the vector potential.

We assume that the dynamical variables take small values, so that we can approximate the Hamiltonian
by expanding the square root to some order in the dynamical variables.  In the conventional paraxial
approximation, the expansion is made to second order.  Here, we expand to third order, and obtain:
\begin{equation}
H^\prime \approx H_{1s} + H_{1y} + H_{1x} + H_2 + H_3 - 1,
\end{equation}
where:
\begin{eqnarray}
H_{1s} & = & p_s + ( k_0 - h ) x + \frac{1}{2}h k_0 x^2, \\
H_{1y} & = & \frac{1}{2} \left( 1 + h x - \frac{\delta}{\beta_0} \right) (p_y - a_y)^2, \\
H_{1x} & = & \frac{1}{2} \left( 1 + h x - \frac{\delta}{\beta_0} \right) (p_x - a_x)^2, \\
H_2 & = & \frac{\phi}{\beta_0}
+ \frac{(\delta - \phi)^2}{2\beta_0^2 \gamma_0^2} \left( 1 + h x - \frac{\delta - \phi}{\beta_0} \right) \nonumber \\
& &  \qquad - \frac{\delta - \phi}{\beta_0} h x, \\
H_3 & = & \frac{\phi}{2\beta_0} \left( (p_x - a_x)^2 + (p_y - a_y)^2 \right) .
\end{eqnarray}
Viewed as a Hamiltonian in its own right, the term $H_{1s}$ is integrable, but this is not the case for the
other terms, $H_{1y}$, $H_{1x}$, $H_2$ or $H_3$.  However, by making appropriate canonical transformations
to new variables, we can express $H_{1y}$, $H_{1x}$ and $H_2$ in integrable form.  $H_3$ is of order 3 (or higher) in the
dynamical variables; we assume we can drop this term (with some loss of accuracy in the solution to the equations
of motion).  We can then construct an explicit symplectic integrator as follows:
\begin{equation}
e^{-\Delta s \, :H:} \approx
e^{-\frac{\Delta s}{2} \, :H_{1}:}
e^{-\Delta s \, :H_2:}
e^{-\frac{\Delta s}{2} \, :H_{1}:},
\end{equation}
where:
\begin{equation}
H_1 = H_{1s} + H_{1y} + H_{1x}.
\end{equation}
Continuing the process:
\begin{equation}
e^{-\frac{\Delta s}{2} \, :H_1:} \approx
e^{-\frac{\Delta s}{4} \, :H_{1s} + H_{1y}:}
e^{-\frac{\Delta s}{2} \, :H_{1x}:}
e^{-\frac{\Delta s}{4} \, :H_{1s} + H_{1y}:},
\end{equation}
and finally:
\begin{equation}
e^{-\frac{\Delta s}{4} \, :H_{1s} + H_{1y}:} \approx
e^{-\frac{\Delta s}{8} \, :H_{1s}:}
e^{-\frac{\Delta s}{4} \, :H_{1y}:}
e^{-\frac{\Delta s}{8} \, :H_{1s}:}.
\end{equation}

The transformations associated with the generator $H_{1s}$ are:
\begin{eqnarray}
e^{-\frac{\Delta s}{8} \, :H_{1s}:} s & = & s + \frac{\Delta s}{8}, \\
e^{-\frac{\Delta s}{8} \, :H_{1s}:} p_x & = & p_x - \frac{\Delta s}{8} (k_0 - h + k_0 h x),
\end{eqnarray}
with the transformations of all other variables (not shown explicitly) corresponding to the identity.

Now consider $H_{1y}$.  To find an explicit form for the transformation generated by $H_{1y}$,
we first consider a transformation to new variables, defined by a mixed-variable generating function:
\begin{equation}
F_y( X_i, p_i; \sigma ) = I_Y - X p_x - Y p_y - Z \delta - S p_s,
\label{mvgff31}
\end{equation}
where $X_i = (X, Y, Z, S)$ are the new co-ordinates,
$p_i = (p_x, p_y, \delta, p_s)$ are the original momenta,
and $I_Y$ is defined by:
\begin{equation}
I_Y = \int_0^Y a_y(X,\bar{Y},S) \, d\bar{Y}.
\end{equation}
In Goldstein's nomenclature \cite{goldstein} $F_y( X_i, p_i; \sigma )$ is a mixed-variable generating
function of the third kind.  The new co-ordinates $(X, Y, Z, S)$ are identical to the original
co-ordinates $(x, y, z, s)$, since:
\begin{equation}
x = - \frac{\partial F_y}{\partial p_x} = X,
\end{equation}
and similarly for $y$, $z$ and $s$.  The new momenta are:
\begin{eqnarray}
P_X & = & - \frac{\partial F_y}{\partial X} = p_x - \frac{\partial I_Y}{\partial X}, \label{newpx1} \\
P_Y & = & - \frac{\partial F_y}{\partial Y} = p_y - a_y, \label{newpy1} \\
P_S & = & - \frac{\partial F_y}{\partial S} = p_s - \frac{\partial I_Y}{\partial S}, \label{newps1}
\end{eqnarray}
and:
\begin{equation}
P_Z = \delta. \label{newpz1}
\end{equation}
In terms of the new variables, $H_{1y}$ can be written:
\begin{equation}
H_{1y} = \frac{1}{2} \left( 1 + h X - \frac{P_Z}{\beta_0} \right) P_Y^2.
\label{h1integrable}
\end{equation}
Viewed as a Hamiltonian, $H_{1y}$ is integrable.  The transformations (generated by $H_{1y}$) of the dynamical variables are:
\begin{eqnarray}
e^{-\frac{\Delta s}{4} \, :H_{1y}:} P_X & = & P_X - \frac{\Delta s}{8} h P_Y^2, \label{newpx2} \\
e^{-\frac{\Delta s}{4} \, :H_{1y}:} Y & = & Y + \frac{\Delta s}{4} \left( 1 + h X - \frac{P_Z}{\beta_0} \right) P_Y, \label{newy1} \\
e^{-\frac{\Delta s}{4} \, :H_{1y}:} Z & = & Z - \frac{\Delta s}{8\beta_0} P_Y^2. \label{newz1}
\end{eqnarray}
Again, the transformations of all other variables (i.e.~for those variables not shown explicitly, above) are given by the identity
transformation.
To apply the transformation $e^{-\frac{\Delta s}{4} \, :H_{1y}:}$, we first transform from the original variables
to a set of new variables using (\ref{newpx1})--(\ref{newps1}); we then apply the transformations
(\ref{newpx2})--(\ref{newz1}), and finally transform back to the original variables using the inverse of the transformations
(\ref{newpx1})--(\ref{newps1}).  Note that although the new momenta do not change under the transformation
generated by $H_{1y}$, the change in the $Y$ co-ordinate leads to a change in $p_x$, $p_y$ and $p_s$ because the
inverse of transformations (\ref{newpx1})--(\ref{newps1}) have to be calculated at a different point from the
original transformations.  Thus:
\begin{eqnarray}
e^{-\frac{\Delta s}{4} \, :H_{1y}:} p_x & = & p_x - \frac{\Delta s}{8} h (p_y - a_y(x,y_0,s))^2 \nonumber \\
& & \qquad + \int_{y_0}^{y_1} \frac{\partial}{\partial x} a_y(x,\bar{y},s)\, d\bar{y}, \label{pxchangegenh1y} \\
e^{-\frac{\Delta s}{4} \, :H_{1y}:} p_y & = & p_y + a_y(x,y_1,s) - a_y(x,y_0,s),
\end{eqnarray}
where $y_0$ and $y_1$ correspond to the initial and final values of the co-ordinate $y$ under the transformation
$e^{-\frac{\Delta s}{4} \, :H_{1y}:}$.  There is also a change in $p_s$; but this has no effect on the dynamics.
In summary, to apply the transformation $e^{-\frac{\Delta s}{4} \, :H_{1y}:}$ we need to evaluate $a_y$ (at
the initial value of the co-ordinate $y = y_0$, and at the final value of the co-ordinate $y = y_1$), and the integral
(with respect to $y$) of the derivative of $a_y$ (with respect to $x$).

In Section~\ref{secvectorpotential} we give analytical expressions for the components of the vector potential, based
on a three-dimensional ``multipole'' decomposition of a magnetic field in a region with a curved reference trajectory.
It is also possible to write down expressions for the derivatives of the vector potential; however, the integral in
(\ref{pxchangegenh1y}) needs to be performed numerically.  Although this will make a significant contribution
to the computational cost for each step in the tracking calculation, in most cases the integral should converge
reasonably quickly given that the derivative of the potential (which is related to the field strength) should vary
slowly over the range of the integral (corresponding to the change in the $y$ co-ordinate over the tracking step).


The transformation with generator $H_{1x}$ may be handled in a similar way to that generated by $H_{1y}$, by
first transforming to new variables.  For the case of $H_{1x}$, we use the mixed-variable generating function:
\begin{equation}
F_x( X_i, p_i; \sigma ) = I_X - X p_x - Y p_y - Z \delta - S p_s,
\label{mvgff3y}
\end{equation}
where:
\begin{equation}
I_X = \int_0^X a_x(\bar{X},Y,S) \, d\bar{X}.
\end{equation}
Note that the new variables in this case (co-ordinates $X$, $Y$, $Z$ and $S$, and momenta $P_X$, $P_Y$, $P_Z$
and $P_S$) are formally different from the variables in the previous case; but to avoid introducing further notation, we use
the same symbols.
The transformations (generated by $H_{1x}$) of the dynamical variables are:
\begin{eqnarray}
e^{-\frac{\Delta s}{2} \, :H_{1x}:} x & = & \frac{\Delta s}{2} \left( 1 - \frac{\delta}{\beta_0} \right) \left( 1 + \frac{\Delta s}{8}h P_X \right) P_X \nonumber \\
& & \qquad + \left( 1 + \frac{\Delta s}{4} h P_X \right)^2 x, \\
e^{-\frac{\Delta s}{2} \, :H_{1x}:} p_x & = & \frac{P_X}{1 + \frac{\Delta s}{4} h P_X} + a_x(x_1,y,s), \\
e^{-\frac{\Delta s}{2} \, :H_{1x}:} p_y & = & p_y + \int_{x_0}^{x_1} \frac{\partial}{\partial y} a_x(\bar{x},y,s) \, d\bar{x},
\label{pychangegenh1x} \\
e^{-\frac{\Delta s}{2} \, :H_{1x}:} z & = & z - \frac{\Delta s}{4\beta_0} \frac{P_X^2}{(1 + \frac{\Delta s}{4}h P_X)},
\end{eqnarray}
where:
\begin{equation}
P_X = p_x - a_x(x_0,y,s),
\end{equation}
and $x_0$ and $x_1$ are the values of $x$ before and after the transformation, respectively.  The variables
$y$ and $\delta$ are unchanged by the transformation.

Finally, we find explicit expressions for the transformation with generator $H_2$ by again first transforming to
new variables.  In this case, we use a mixed-variable generating function:
\begin{equation}
F_3^\prime( X_i^\prime, p_i; \sigma ) = \phi(X^\prime, Y^\prime, S^\prime) Z^\prime - X^\prime p_x - Y^\prime p_y - Z^\prime \delta - S^\prime p_s,
\label{mvgff3}
\end{equation}
where $X_i^\prime = (X^\prime, Y^\prime, Z^\prime, S^\prime)$ are the new co-ordinates, and
$p_i = (p_x, p_y, \delta, p_s)$ are the original momenta.  The new co-ordinates are identical to the original
co-ordinates, since:
\begin{equation}
x = - \frac{\partial F_3^\prime}{\partial p_x} = X^\prime,
\end{equation}
and similarly for $y$, $z$ and $s$.  The new momenta are:
\begin{eqnarray}
P_X^\prime & = & - \frac{\partial F_3^\prime}{\partial X^\prime} = p_x - \frac{\partial \phi}{\partial X^\prime} Z^\prime, \label{newpx} \\
P_Y^\prime & = & - \frac{\partial F_3^\prime}{\partial Y^\prime} = p_y - \frac{\partial \phi}{\partial Y^\prime} Z^\prime, \\
P_S^\prime & = & - \frac{\partial F_3^\prime}{\partial S^\prime} = p_s - \frac{\partial \phi}{\partial S^\prime} Z^\prime,
\end{eqnarray}
and:
\begin{equation}
P_Z^\prime = \delta - \phi. \label{newpz}
\end{equation}
In terms of the new variables, $H_2$ can be written:
\begin{equation}
H_2 = \frac{\phi}{\beta_0}
+ \frac{P_Z^{\prime 2}}{2\beta_0^2 \gamma_0^2} \left( 1 + h X^\prime - \frac{P_Z^\prime}{\beta_0} \right)
- \frac{h}{\beta_0} X^\prime P_Z^\prime, \label{h3newvars}
\end{equation}
which is an integrable Hamiltonian,
leading to the transformations:
\begin{eqnarray}
e^{-\Delta s \, :H_2:} P_X^\prime & = & P_X^\prime - \frac{\Delta s}{\beta_0} \frac{\partial \phi}{\partial X^\prime}
- \Delta s \, \frac{h P_Z^{\prime 2}}{2\beta_0^2 \gamma_0^2} + \Delta s \frac{h}{\beta_0} P_Z^\prime, \nonumber \\
& & \\
e^{-\Delta s \, :H_2:} P_Y^\prime & = & P_Y^\prime - \frac{\Delta s}{\beta_0} \frac{\partial \phi}{\partial Y^\prime}, \\
e^{-\Delta s \, :H_2:} P_S^\prime & = & P_S^\prime  - \frac{\Delta s}{\beta_0} \frac{\partial \phi}{\partial S^\prime}, \\
e^{-\Delta s \, :H_2:} Z^\prime & = & Z^\prime - \frac{\Delta s}{\beta_0}h X^\prime \nonumber \\
& & \qquad + \Delta s \, \frac{P_Z^\prime}{\beta_0^2 \gamma_0^2} \,\left( 1 + h X^\prime - \frac{3P_Z^\prime}{2\beta_0} \right).
\end{eqnarray}
Again, transformations of the variables not given explicitly above, are equal to the identity. 

\section{$s$-dependent fields in toroidal co-ordinates}
Applying the symplectic integrator described in Section~\ref{sectionderivation} involves derivatives of the
scalar potential, and derivatives and integrals of the vector potential.  It is therefore helpful to have analytic
representations of the scalar and vector potentials, from which expressions for the derivatives and integrals
may be found.  In practice, however, only a purely numerical representation of the potentials may be
available (giving, for example, the values of the potentials on a grid of discrete points over some region of
space).  With a straight reference trajectory ($h = 0$), it is possible to fit the coefficients of series
representations of the potentials, for example using generalised gradients \cite{dragt}; the series
representation gives the functional dependence of the potential on the co-ordinates, and this therefore
provides a suitable representation for applying the integrator.


A similar approach is possible in the case that the reference trajectory has some non-zero curvature.
Expressions for ``curvilinear multipoles'' (multipole fields around curved reference trajectories) have been
given by McMillan and others \cite{mcmillan,zolkin,schnizer,brouwer}, and have been implemented in the tracking
code Bmad \cite{bmad}.  However, the available expressions are not ideal for use where the potential is
given in purely numerical form.  In much of the previous work, the multipoles are expressed in terms of
the transverse Cartesian co-ordinates, $x$ and $y$: obtaining the multipole coefficients then involves
fitting polynomials to the numerical data along either the $x$ or $y$ axis \cite{herrod2017}.  The nature of the potential
(which satisfies Laplace's equation) is such that residuals to the fit will grow exponentially with distance
from the line along which the fit is performed.  A more robust approach is based on fitting to a surface
bounding some region of space enclosing the reference trajectory: within the surface, the residuals
decrease exponentially with distance from the surface.  Although the residuals will still grow exponentially
outside the region enclosed by the surface, if the surface is chosen appropriately then the enclosed region
will cover the volume of interest for particle tracking.

\begin{figure}
\begin{center}
\includegraphics[width=\columnwidth]{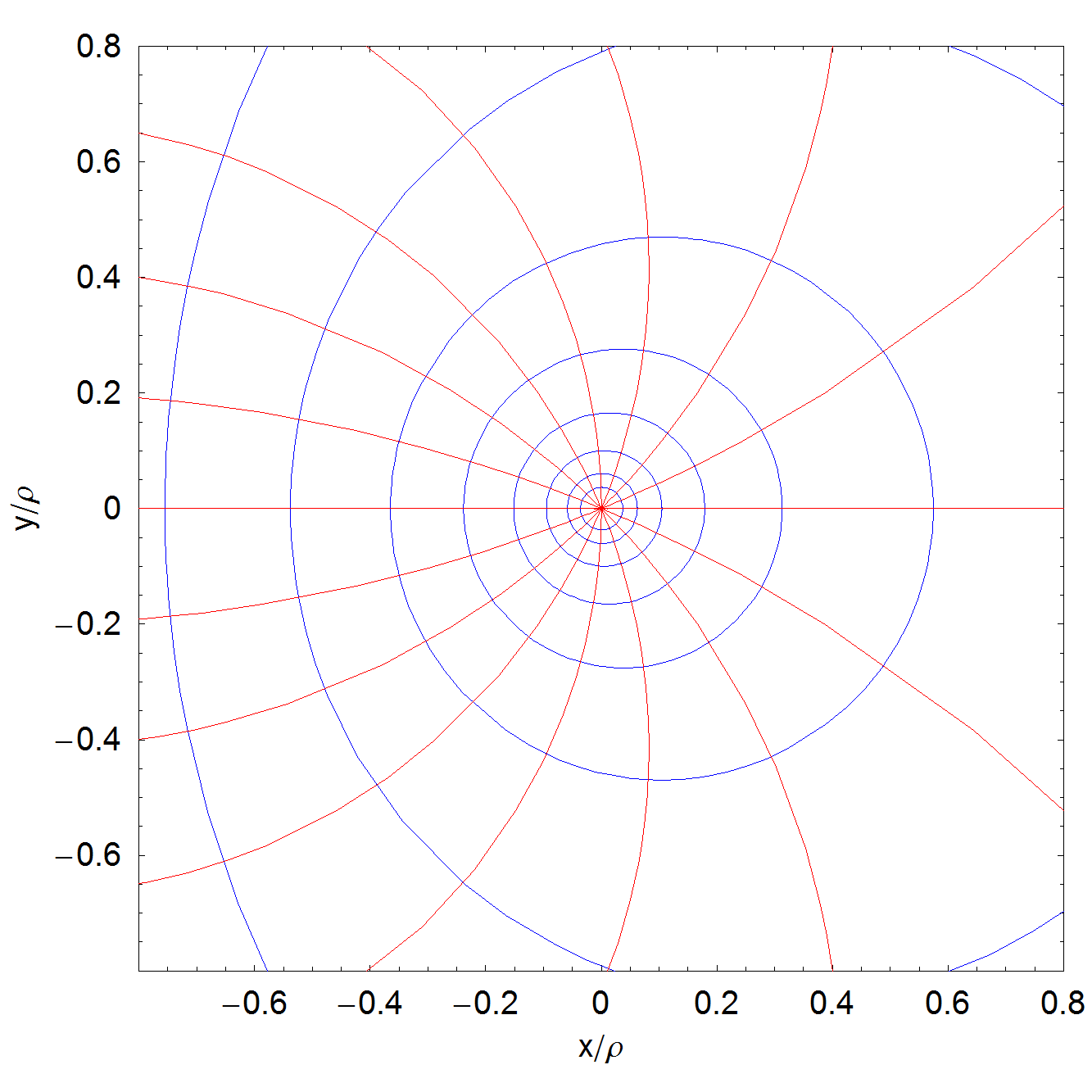}
\caption{Toroidal co-ordinates.  The red curves show lines of constant $v$ from 0 to 2$\pi$.
The blue curves show lines of constant value for the co-ordinate $u$ in the range 0.5 to 4 in steps of 0.5,
with $0 \leq v < 2\pi$.  Larger values of $u$ give circles of smaller diameter; in the limit $u \to \infty$, the
circles converge towards the reference trajectory $x = y = 0$.
\label{figtoroidalcoordinates}}
\end{center}
\end{figure}

To obtain a multipole decomposition based on fitting numerical data on a surface, it is convenient in the
case of a curved reference trajectory to work in toroidal co-ordinates \cite{arfken,mathworldtoroidal}.
The co-ordinates in the transverse plane are illustrated in Fig.~\ref{figtoroidalcoordinates}.  The toroidal
co-ordinates $u$ and $v$ are related to the accelerator co-ordinates $x$ and $y$ (Cartesian co-ordinates
in a plane perpendicular to the reference trajectory) by:
\begin{eqnarray}
x & = & \rho \left( \frac{\sinh(u)}{\cosh(u) - \cos(v)} - 1 \right), \\
y & = & \frac{\rho \sin(v)}{\cosh(u) - \cos(v)},
\end{eqnarray}
where $\rho = 1/h$ is the radius of curvature of the reference trajectory.  The longitudinal co-ordinate
$s$ (the distance along the reference trajectory) is related to the toroidal co-ordinate $\theta$ by:
\begin{equation}
s = \rho \theta.
\end{equation}
A surface enclosing the reference trajectory can be defined by specifying a fixed value $u_\mathrm{ref}$
for the co-ordinate $u$: a surface defined by $u = u_\mathrm{ref}$ for $0 \le v < 2\pi$ and $0 \le \theta < 2\pi$
resembles a torus.  If numerical field data are available for the scalar and vector potentials on such
a surface, then it is possible to fit the coefficients of series expansions for the scalar and vector
potentials (up to some desired order) to the data.  This produces expressions that are suitable for
use in the explicit symplectic integrator described in Section~\ref{sectionderivation}.  We first discuss
the case of the scalar potential, and then extend the results to the vector potential.

\subsection{Scalar potential in toroidal co-ordinates}
In terms of the toroidal co-ordinates, an harmonic potential (such that $\nabla^2 \phi = 0$) may be
written \cite{arfken,mathworldlaplacian}:
\begin{equation}
\phi = \sum_{m,n=-\infty}^\infty f_{mn} \phi_{mn},
\label{toroidalpotential}
\end{equation}
where the $f_{mn}$ are coefficients representing the strength of a multipole component $\phi_{mn}$.
The multipole components are given by:
\begin{equation}
\phi_{mn} = (-i)^m \mathcal{C}(u,v) \, P_{n-\frac{1}{2}}^{-|m|}(\coth(u)) \, e^{i mv}e^{i n\theta},
\label{phimn}
\end{equation}
where $P_\nu^\mu(\xi)$ is an associated Legendre polynomial of the first kind, and:
\begin{equation}
\mathcal{C}(u,v) = \sqrt{\frac{\cosh(u)-\cos(v)}{\sinh(u)}} = \sqrt{\frac{\rho}{x + \rho}}.
\end{equation}
An algorithm for computation of the associated Legendre polynomials with positive $\mu$ has been
presented by Segura and Gil \cite{seguragil};  values for negative $\mu$ are readily obtained using \cite{stegun}:
\begin{eqnarray}
P_\nu^{-\mu}(\xi) & = & \frac{\Gamma(\nu-\mu+1)}{\Gamma(\nu+\mu+1)} \nonumber \\
& & \times \left( P_\nu^\mu(\xi) - \frac{2}{\pi}e^{-i\mu\pi}\sin(\mu\pi)Q_\nu^\mu(\xi) \right),
\label{pnegativem}
\end{eqnarray}
where $Q_\nu^\mu(\xi)$ is an associated Legendre polynomial of the second kind.  Note that for
integer $\mu$ (which is the case of interest here), the term in $Q_\nu^\mu(\xi)$ in (\ref{pnegativem}) vanishes.

We shall show in Section~\ref{sectoroidalpotentials} that each component $\phi_{mn}$ has properties
that may be expected of a multipole of order $m$, with $m=1$ corresponding to a dipole, $m=2$ a
quadrupole, and so on.  Note that a normal dipole deflects a particle horizontally, whereas a skew dipole
deflects a particle vertically.

Given numerical data for a potential $\phi(u,v,\theta)$, the coefficients $f_{mn}$ may be obtained from:
\begin{equation}
f_{mn} = \frac{1}{N_{mn}} \int_0^{2\pi} dv \int_0^{2\pi} d\theta \,
\frac{e^{-imv}e^{-in\theta}\phi(u_\mathrm{ref},v,\theta)}{\sqrt{\cosh(u_\mathrm{ref}) - \cos(v)}} ,
\end{equation}
where $u_\mathrm{ref}$ is a fixed value of $u$ that defines the surface (enclosing the reference trajctory,
$x = y = 0$) on which the fit to the numerical data is performed, and $N_{mn}$ is a normalising factor:
\begin{equation}
N_{mn} = (-i)^m 4\pi^2 \frac{P_{n-\frac{1}{2}}^{-|m|}(\coth(u_\mathrm{ref}))}{\sqrt{\sinh(u_\mathrm{ref})}} .
\end{equation}

As an alternative to calculating the coefficients $f_{mn}$ from the scalar potential, they may be calculated
from the electric field components.  The electric field is derived from the potential by:
\begin{equation}
\mathbf{E} = (E_u, E_v, E_\theta) = - \nabla \Phi = - \frac{cP_0}{q}\nabla \phi.
\end{equation}
The $E_v$ component of the field (tangential to a line defined by fixed values of $u$ and $\theta$) is given by:
\begin{widetext}
\begin{equation}
E_v = - \frac{cP_0}{q} \sum_{m,n} f_{mn} \frac{(-i)^m}{\rho}
\left( \frac{1}{2} \sin(v) + im (\cosh(u) -\cos(v)) \right)
\mathcal{C}(u,v) P_{n-\frac{1}{2}}^{-|m|}(\coth(u)) e^{imv} e^{in\theta}.
\end{equation}
The coefficients $f_{mn}$ can then be found from the values of $E_v$ on a surface $u = u_\mathrm{ref}$:
\begin{equation}
f_{mn} = \frac{1}{N^\prime_{mn}} \int_0^{2\pi} dv \int_0^{2\pi} d\theta
\frac{e^{-imv} e^{-in\theta} E_v(u_\mathrm{ref},v,\theta)}{\left( \frac{1}{2} \sin(v) + im (\cosh(u_\mathrm{ref}) -\cos(v)) \right) \mathcal{C}(u_\mathrm{ref},v)},
\end{equation}
\end{widetext}
where:
\begin{equation}
N^\prime_{mn} = -\frac{(-i)^m}{\rho} 4\pi^2 P_{n-\frac{1}{2}}^{-|m|}(\coth(u_\mathrm{ref})).
\end{equation}

To apply the symplectic integrator described in Section~\ref{sectionderivation}, we need
the derivatives of the potential with respect to the Cartesian co-ordinates.  The derivates
can be obtained from:
\begin{eqnarray}
\frac{\partial \phi}{\partial x} & = &
\frac{\partial \phi}{\partial u} \frac{\partial u}{\partial x} +
\frac{\partial \phi}{\partial v} \frac{\partial v}{\partial x},
\label{dphidx} \\
\frac{\partial \phi}{\partial y} & = &
\frac{\partial \phi}{\partial u} \frac{\partial u}{\partial y} +
\frac{\partial \phi}{\partial v} \frac{\partial v}{\partial y},
\label{dphidy}
\end{eqnarray}
and:
\begin{equation}
\frac{\partial \phi}{\partial s} = \frac{\partial \phi}{\partial \theta} \frac{\partial \theta}{\partial s}
= \frac{1}{\rho}\frac{\partial \phi}{\partial \theta}.
\end{equation}
For a given multipole component (\ref{phimn})
the derivatives with respect to the toroidal co-ordinates $u$ and $v$ are:
\begin{eqnarray}
\frac{\partial \phi_{mn}}{\partial u} & = & (-i)^m 
\left( \left( n\coth(u) + \frac{1}{2\mathcal{C}(u,v)} \right) P_{n-\frac{1}{2}}^{-|m|}(\coth(u)) \right. \nonumber \\
& & - \left. \left( |m| + n + \frac{1}{2} \right) \mathcal{C}(u,v) P_{n+\frac{1}{2}}^{-|m|}(\coth(u)) \right) \nonumber \\
& & \times e^{imv} e^{in\theta}, \label{dphindu}
\end{eqnarray}
and:
\begin{eqnarray}
\frac{\partial \phi_{mn}}{\partial v} & = & (-i)^m \left( \frac{\sin(v)}{2\sinh(u)\mathcal{C}(u,v)} + im\mathcal{C}(u,v) \right)
\nonumber \\
& & \times P_{n-\frac{1}{2}}^{-|m|}(\coth(u)) e^{imv} e^{in\theta}. \label{dphindv}
\end{eqnarray}
Finally, we need the derivatives of the toroidal co-ordinates $(u,v)$ with respect to the
Cartesian co-ordinates $(x,y)$.  The toroidal co-ordinates can be expressed in terms of the
Cartesian co-ordinates as follows:
\begin{equation}
u - i v = 2\coth^{-1}\! \left(1 + \frac{x + iy}{\rho}\right).
\end{equation}
We then find:
\begin{eqnarray}
\frac{\partial u}{\partial x} = \frac{\partial v}{\partial y} & = &
\frac{-2\rho \big( (2\rho + x)x - y^2 \big)}{(x^2 + y^2) \big( (2\rho + x)^2 + y^2 \big)} \nonumber \\
& = & \frac{1}{\rho}(1 - \cosh(u) \cos(v)),
\label{dudx}
\end{eqnarray}
and:
\begin{eqnarray}
\frac{\partial u}{\partial y} = -\frac{\partial v}{\partial x} & = &
\frac{-4\rho (\rho + x)y}{(x^2 + y^2) \big( (2\rho + x)^2 + y^2 \big)} \nonumber \\
& = & \frac{1}{\rho}\sinh(u) \sin(v).
\label{dudy}
\end{eqnarray}
The derivatives of the potential with respect to the Cartesian co-ordinates can be found by
using equations (\ref{dphindu}), (\ref{dphindv}), (\ref{dudx}) and (\ref{dudy}) in
equations (\ref{dphidx}) and (\ref{dphidy}).  Tracking a particle through a field
described by a scalar potential can then be achieved by using the potential and its derivatives
(with respect to $x$ and $y$) in the symplectic integrator described in Section~\ref{sectionderivation}.


\subsection{Vector potential in toroidal co-ordinates\label{secvectorpotential}}
To apply the explicit symplectic integrator to a particle moving through a magnetic field, we need expressions
for the components of the vector potential.  Since we address the case of a curved reference trajectory,
we assume that the magnetic field has a (normal) dipole component derived from the longitudinal component
$a_s$ of the vector potential (\ref{magneticvectorpotential}).  Other components of the magnetic field
(corresponding to quadrupole, or higher-order multipole components) may be derived from the transverse
components of the vector potential.  In toroidal co-ordinates, these components may be expressed as
follows:
\begin{eqnarray}
a_u & = & i \sinh(u) \sum_{m,n=-\infty}^\infty \frac{\alpha_{mn}}{n} \frac{\partial \phi_{mn}}{\partial v}, \label{vectorpotentialu} \\
a_v & = & -i \sinh(u) \sum_{m,n=-\infty}^\infty \frac{\alpha_{mn}}{n} \frac{\partial \phi_{mn}}{\partial u}, \label{vectorpotentialv}
\end{eqnarray}
where the functions $\phi_{mn}$ are given by (\ref{phimn}).
In the case that $a_\theta = 0$ (i.e.~the longitudinal component of the vector potential is zero, so that
$k_0 = 0$ in (\ref{magneticvectorpotential})), and $\alpha_{mn} = f_{mn}$ for all $m$, $n$, it is found that:
\begin{equation}
\nabla \times \mathbf{a} = -\nabla \phi,
\end{equation}
with $\phi$ given by (\ref{toroidalpotential}).  Hence, the magnetic field derived from the vector potential
$\mathbf{a} = (a_u, a_v, 0)$ with components (in toroidal co-ordinates) given by (\ref{vectorpotentialu}) and
(\ref{vectorpotentialv}) has the same form as the electric field derived from the scalar potential $\phi$ given by
(\ref{toroidalpotential}).

To apply the symplectic integrator described in Section~\ref{sectionderivation}, we require the components of
the vector potential in Cartesian co-ordinates, and their derivatives.  Given the components $(a_u, a_v)$ in toroidal co-ordinates,
the components $(a_x, a_y)$ in Cartesian co-ordinates are obtained from:
\begin{widetext}
\begin{eqnarray}
a_x & = & \frac{1}{N} \frac{\partial x}{\partial u} a_u + \frac{1}{N} \frac{\partial x}{\partial v} a_v
 = \frac{(1 - \cosh(u) \cos(v))a_u - \sinh(u)\sin(v) a_v}{\cosh(u) - \cos(v)}, \\
a_y & = & \frac{1}{N} \frac{\partial y}{\partial u} a_u + \frac{1}{N} \frac{\partial y}{\partial v} a_v
 = -\frac{(1 - \cosh(u) \cos(v))a_v + \sinh(u)\sin(v) a_u}{\cosh(u) - \cos(v)}, 
\end{eqnarray}
\end{widetext}
where the normalising factor $N$ is:
\begin{eqnarray}
N & = & \sqrt{ \left( \frac{\partial x}{\partial u} \right)^2 + \left( \frac{\partial y}{\partial u} \right)^2 }
=\sqrt{ \left( \frac{\partial x}{\partial v} \right)^2 + \left( \frac{\partial y}{\partial v} \right)^2 } \nonumber \\
& = & \frac{\rho}{\cosh(u) - \cos(v)}.
\end{eqnarray}
The derivatives of $a_x$ and $a_y$ with respect to the Cartesian co-ordinates $x$ and $y$ can be expressed
in terms of the derivatives with respect to the toroidal co-ordinates $u$ and $v$:
\begin{eqnarray}
\frac{\partial a_x}{\partial y} & = & \frac{\partial u}{\partial y} \frac{\partial a_x}{\partial u} + \frac{\partial v}{\partial y} \frac{\partial a_x}{\partial v}, \\
\frac{\partial a_y}{\partial x} & = & \frac{\partial u}{\partial x} \frac{\partial a_y}{\partial u} + \frac{\partial v}{\partial x} \frac{\partial a_y}{\partial v}.
\end{eqnarray}
Given (\ref{vectorpotentialu}) and (\ref{vectorpotentialv}), the derivatives of $a_x$ and $a_y$ with respect
to the toroidal co-ordinates may be found from the second derivatives of the scalar potential:
\begin{widetext}
\begin{eqnarray}
\frac{\partial^2 \phi_{mn}}{\partial u^2} & = & (-i)^m \left(
\frac{ c_1 }{16\sinh^4(u)\mathcal{C}(u,v)^3} P_{n-\frac{1}{2}}^{-|m|}(\coth(u)) + \right.
\frac{c_2}{\sinh^2(u)\mathcal{C}(u,v)} \left( |m|+n+\frac{1}{2}\right) P_{n+\frac{1}{2}}^{-|m|}(\coth(u)) \nonumber \\
& & \quad + \left. \mathcal{C}(u,v) \left( |m|+n+\frac{1}{2}\right) \left( |m|+n+\frac{3}{2}\right) P_{n+\frac{3}{2}}^{-|m|}(\coth(u))
\right) e^{imv} e^{in\theta}, \\
\frac{\partial^2 \phi_{mn}}{\partial u \, \partial v} & = & (-i)^{m-1} \left(
\frac{ c_3 }{4\sinh^3(u)\mathcal{C}(u,v)^3} P_{n-\frac{1}{2}}^{-|m|}(\coth(u)) +
\frac{c_4}{2\sinh(u)\mathcal{C}(u,v)} \left( |m|+n+\frac{1}{2}\right) P_{n+\frac{1}{2}}^{-|m|}(\coth(u)) \right) e^{imv} e^{in\theta}, \nonumber \\
& & \\
\frac{\partial^2 \phi_{mn}}{\partial v^2} & = & (-i)^m
\frac{ c_5 }{\sinh(u)\mathcal{C}(u,v)} P_{n-\frac{1}{2}}^{-|m|}(\coth(u)) e^{imv} e^{in\theta}.
\end{eqnarray}
where:
\begin{eqnarray}
c_1 & = & 4(1 - 2n(2n-5) - (1+2n+4n^2)\cosh(2u))\cos(v)\cosh(u)
+ 4n(6(n-1) + (n\cosh(2u)+n-2)\cos(2v)) \nonumber \\
& & \quad + (5+4n(7n-3))\sinh(u)^2 + (1+2n)^2\sinh(u)\sinh(3u), \\
c_2 & = & 1+(1+2n)\cos(v)\cosh(u)-2(1+n)\cosh^2(u), \\
c_3 & = & 2m\cos(v) - i\sin(v) + i n \cosh(u) \sin(2v) + 
m (4n-1+2n\cos(2v) + (1+2n)\cosh(2u) ) \cosh(u) \nonumber \\
& & \quad + (i(1-2n)\sin(v) - 2m(1+4n)\cos(v) )\cosh^2(u) , \\
c_4 & = & i\sin(v) - 2m(\cosh(u) - \cos(v)), \\
c_5 & = & \frac{1}{2}\cos(v) - m^2 (\cosh(u) - \cos(v)) + \left( im - \frac{\sin(v)}{4(\cosh(u) - \cos(v))} \right) \sin(v).
\end{eqnarray}
\end{widetext}

\subsection{Examples of multipole potentials in toroidal co-ordinates\label{sectoroidalpotentials}}

To illustrate the scalar potential given by (\ref{toroidalpotential}), we consider the case that the potential is independent
of the longitudinal co-ordinate, $\theta$: as a consequence, we need to include only a single longitudinal mode, $n = 0$ in the
summation in (\ref{toroidalpotential}).  With a straight reference trajectory ($h = 0$), we expect a multipole potential
to take the form:
\begin{equation}
\phi_m = \mathrm{Re}\left( C_m (x + iy)^m \right),
\label{multipolestraightreferencetrajectory}
\end{equation}
where the real and imaginary parts of the coefficient $C_m$ determine the strengths of the normal and skew
components of the field.  Hence, in a normal multipole field of order $m$ the potential varies along the $x$ axis as:
\begin{equation}
\phi_m = \mathrm{Re}(C_m) x^m,
\end{equation}
and along the $y$ axis as:
\begin{equation}
\phi_m = \left\{ \begin{array}{ll} 
\mathrm{Im}(C_m) y^m & \textrm{odd }m, \\
\mathrm{Re}(C_m) y^m & \textrm{even }m.
\end{array} \right.
\end{equation}

\begin{figure*}
\begin{center}
\includegraphics[width=\textwidth]{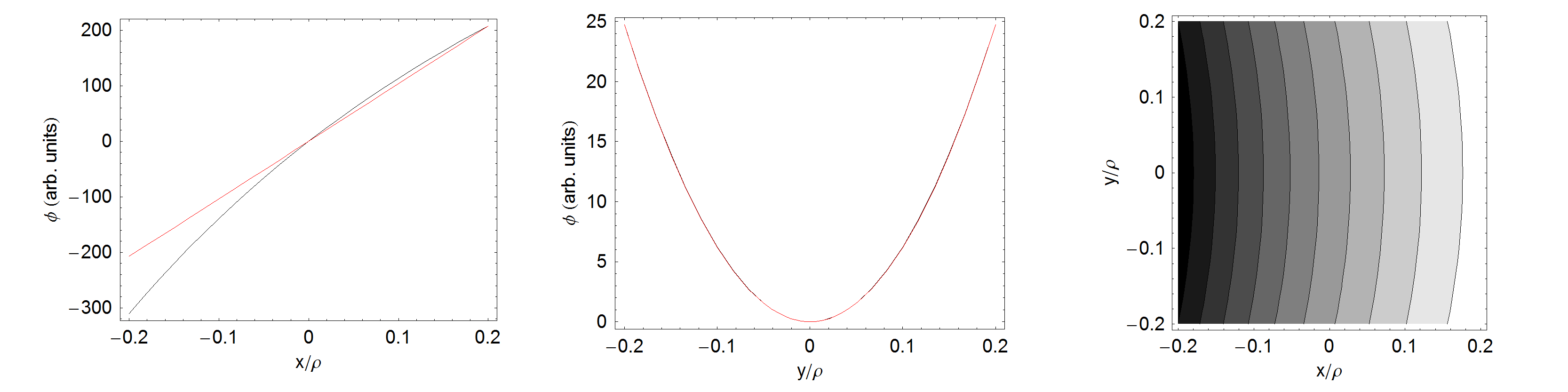}
\includegraphics[width=\textwidth]{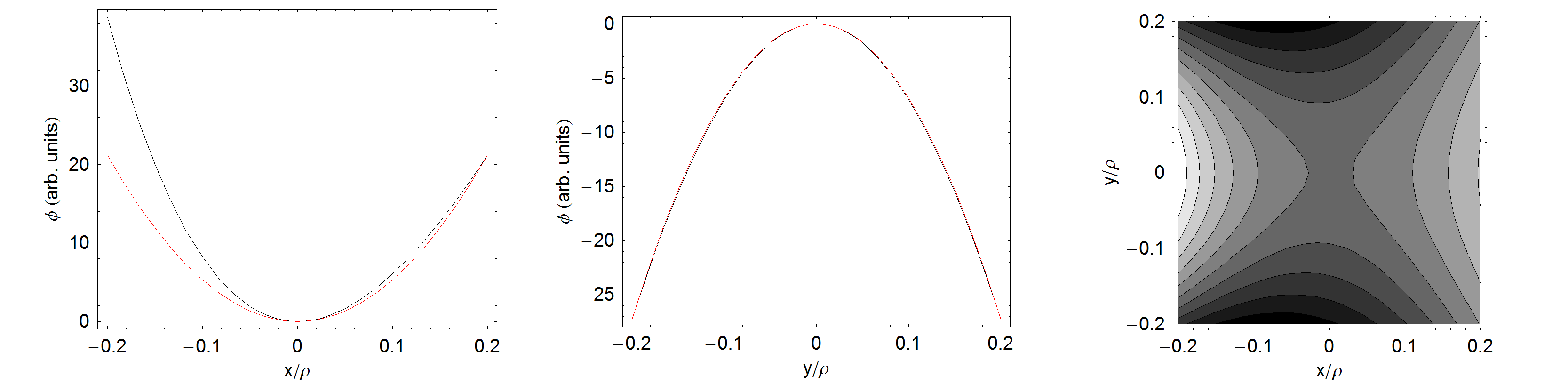}
\includegraphics[width=\textwidth]{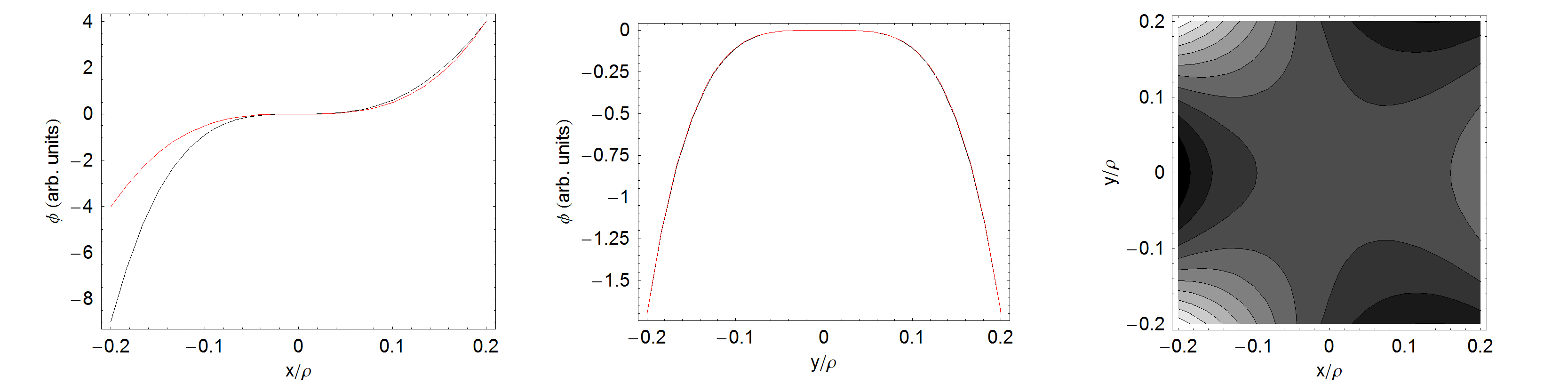}
\caption{Scalar potential in normal multipoles with a curved reference trajectory.
Each row shows (top to bottom) the potential in a multipole of order $n=1$ (dipole), order $n=2$ (quadrupole)
and order $n=3$ (sextupole).
The left-hand and middle plots in each row show respectively the potential (black line) as a function of horizontal position $x$,
with $y = 0$, and as a function of vertical position $y$, with $x = 0$.
The red lines in the left-hand plots show curves $\phi \propto x^n$.
The red lines in the middle plots show curves $\phi \propto y^{n+1}$ for odd $n$, and $\phi \propto y^n$ for even $n$.
The right-hand plot in each row shows contours of constant potential in the plane perpendicular
to the reference trajectory.
\label{fignormamultipolepotential}}
\end{center}
\end{figure*}

\begin{figure*}
\begin{center}
\includegraphics[width=\textwidth]{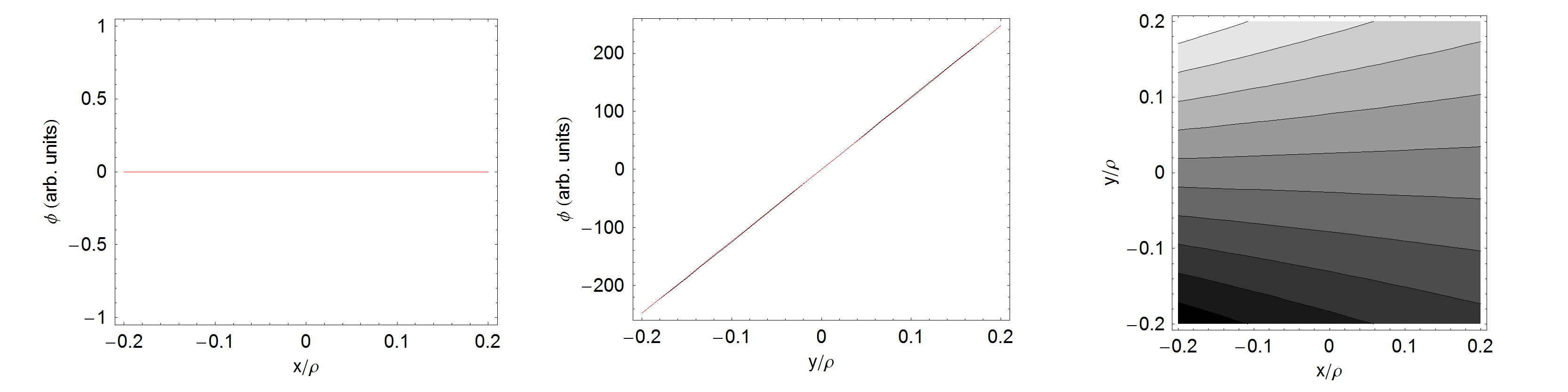}
\includegraphics[width=\textwidth]{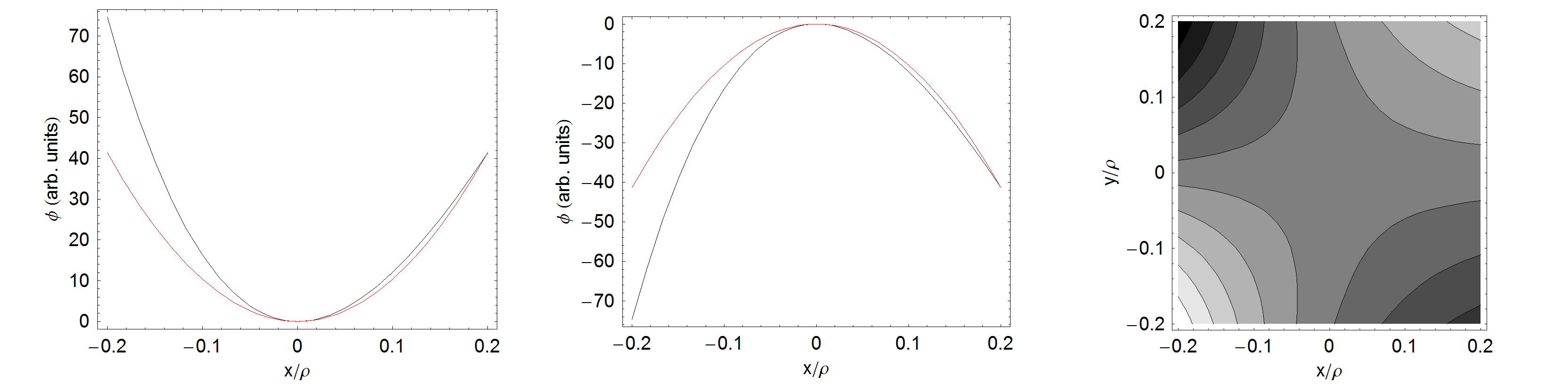}
\includegraphics[width=\textwidth]{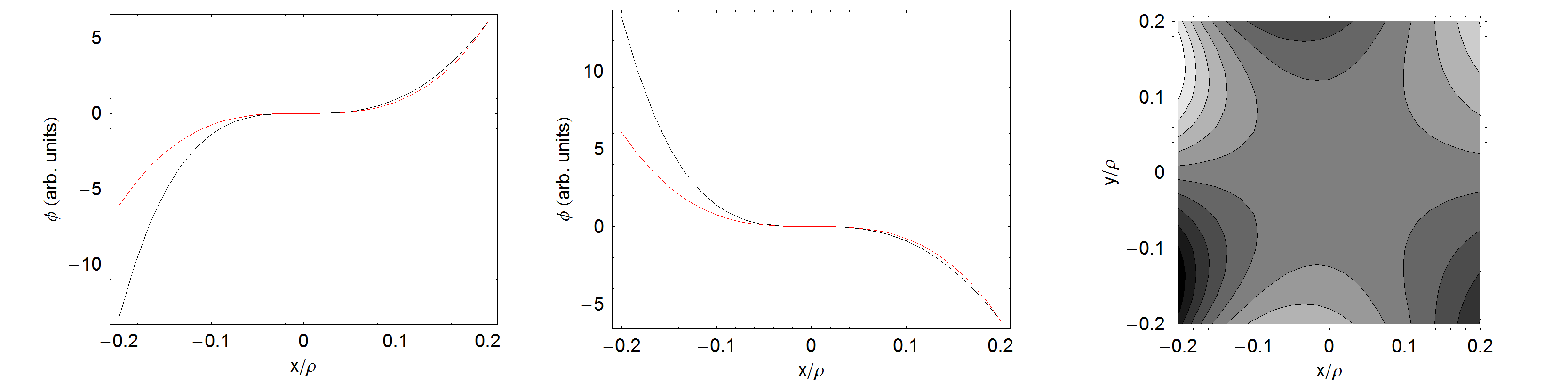}
\caption{Scalar potential in skew multipoles with a curved reference trajectory.
Each row shows (top to bottom) the potential in a multipole of order $n=1$ (dipole), order $n=2$ (quadrupole)
and order $n=3$ (sextupole).
The left-hand and middle plots in the top row (dipole) show respectively the potential as a function of $x$, with $y=0$,
and as a function of $y$, with $x=0$ (black line).
In the middle and bottom rows (quadrupole and sextupole), the left-hand and middle plots show respectively the potential
as a function of $x$, with $y=x\tan(\pi/2n)$, and as a function of $x$, with $y=-x\tan(\pi/2n)$ (black lines).
The red lines in the left-hand and middle plots show curves $\phi \propto x^n$ (or $\phi \propto y^n$ in the top row, middle plot).
The right-hand plot in each row shows contours of constant potential in the plane perpendicular
to the reference trajectory.
\label{figskewmultipolepotential}}
\end{center}
\end{figure*}

With a curved reference trajectory, we expect to see similar behaviour in the dependence of the potential
for a given order of multipole on the $x$ and $y$ co-ordinates, but with some difference from the
dependence given in (\ref{multipolestraightreferencetrajectory}) arising from the curvature.  One way to
show a similarity between multipoles with straight and curved reference trajectories would be to expand
the potential in the case of a multipole with curved reference trajectory as a series in $x$ and $y$; unfortunately,
the fact that the limit $x \to 0$, $y \to 0$ corresponds to $u \to \infty$ makes it problematic to obtain
the appropriate series.  However, we can plot the potential for a given order of (normal or skew) multipole as a
function of $x$ and $y$: plots for dipoles, quadrupoles and sextupoles are shown in
Fig.~\ref{fignormamultipolepotential} (normal multipoles) and Fig.~\ref{figskewmultipolepotential} (skew multipoles).

From Fig.~\ref{fignormamultipolepotential} (top), for example, we see that for a normal dipole the potential has an approximately
linear dependence on $x$.  With a straight reference trajectory, we would expect the potential to be independent
of $y$; however, the curvature of the reference trajectory introduces a second-order dependence of the potential
on $y$.  In the case of a normal quadrupole (Fig.~\ref{fignormamultipolepotential}, middle), the potential has a (roughly)
quadratic dependence on both $x$ and $y$: this again corresponds to the behaviour that we would expect in the
case of a straight reference trajectory.  Because the curvature of the reference trajectory breaks the symmetry
between positive and negative values of $x$, the effect of the curvature is more evident in the dependence of the
potential on $x$, than in the dependence of the potential on $y$.  For a skew quadrupole (Fig.~\ref{figskewmultipolepotential}, middle),
the potential with a straight reference trajectory is exactly zero along the $x$ and $y$ axes.  With a curved
reference trajectory, the potential is zero along the $x$ axis (as required by symmetry); but there is a relatively
weak fourth-order dependence of the potential on $y$ (with $x = 0$).  Other cases demonstrate the general behaviour
we would expect for a multipole potential in a straight co-ordinate system, but with some differences arising from
the curvature of the reference trajectory.

\section{Test cases}

To illustrate application of the explicit symplectic integrator presented in Section~\ref{sectionderivation}, we
consider three test cases: a curvilinear magnetic skew sextupole, a curvilinear electrostatic quadrupole, and the
fringe field region of an electrostatic quadrupole in the g-2 storage ring
\cite{gminus2,gminus2tdrquads,wu2017,semertzidis2003}.  The first two cases
are ``artificial'' in the sense that they are based on fields described by a small number of components; the third case
is more realistic, and uses field component coefficients fitted to numerical data obtained from a modelling code.
In each case, we track a particle with some chosen initial conditions through the field using the explicit symplectic
integrator.  For comparison, we also integrate numerically the (Hamiltonian) equations of motion derived from the exact
Hamiltonian (\ref{hamiltonian}).  All calculations are performed in Mathematica 5.0 \cite{mathematica}; for numerical
integration of the equations of motion derived from the Hamiltonian (\ref{hamiltonian}), we use the NDSolve
function with default settings; although this provides a non-symplectic integration, it should achieve good accuracy.

\begin{figure*}
\begin{center}
\includegraphics[width=0.4\textwidth]{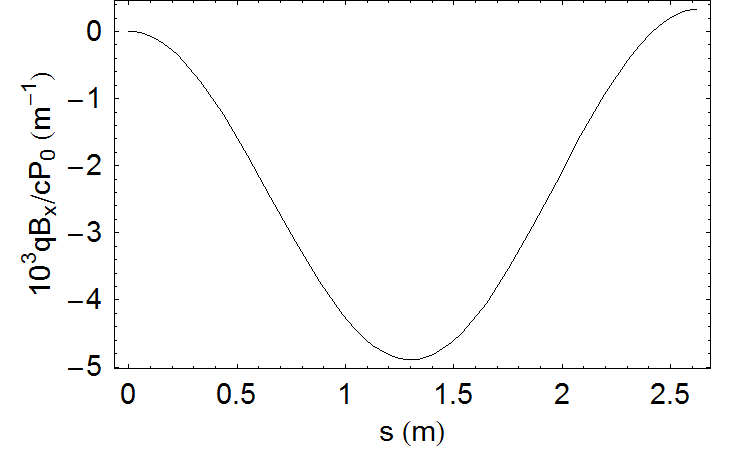}
\includegraphics[width=0.4\textwidth]{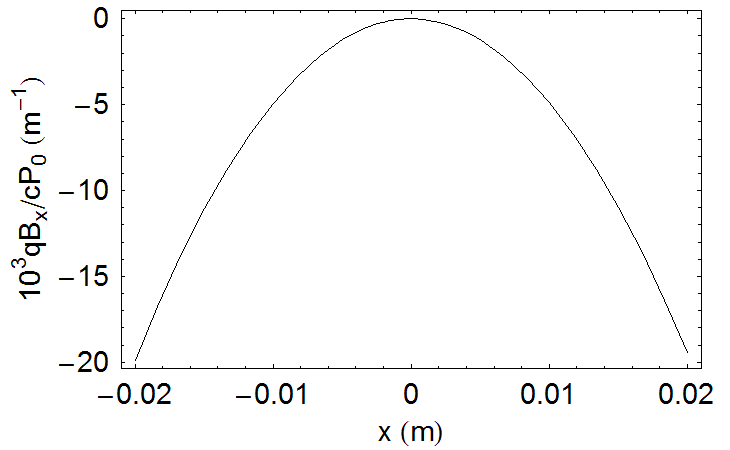} \\
\includegraphics[width=0.4\textwidth]{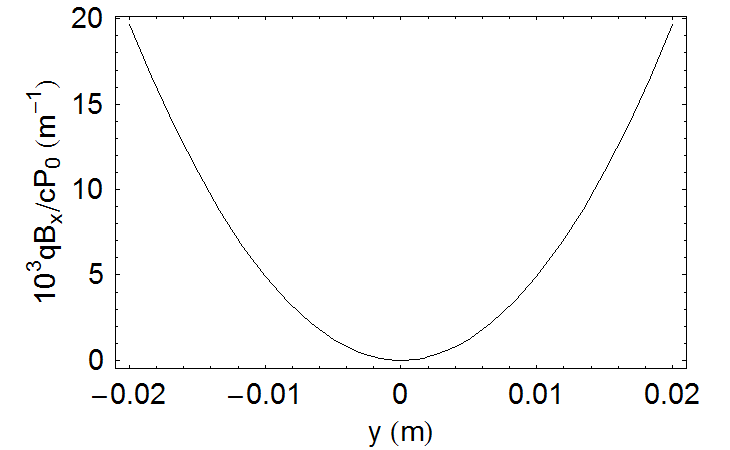}
\includegraphics[width=0.4\textwidth]{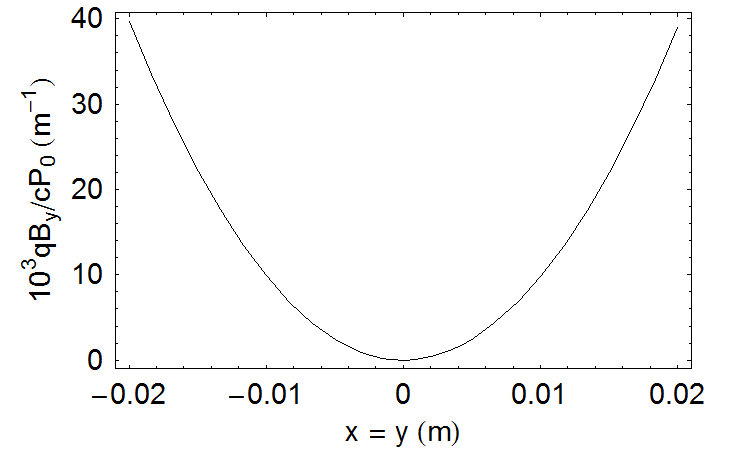}
\caption{Magnetic field in a curvilinear skew sextupole, derived from the scalar potential (\ref{testcasemagpotential}).
Top left: $B_x$ as a function of $s$ for $x = 10\,$mm and $y = 0$.
Top right: $B_x$ as a function of $x$ for $y = 0$ and $s = \frac{\pi}{12}\rho$.
Bottom left: $B_x$ as a function of $y$ for $x = 0$, $s = \frac{\pi}{12}\rho$.
Bottom right: $B_y$ along the line $x = y$, for $s = \frac{\pi}{12}\rho$.
In each plot, the field is scaled by the beam rigidity.
\label{figtestcasemagneticfield}}
\end{center}
\end{figure*}

\subsection{Curvilinear magnetic skew sextupole}
As a first illustration of the explicit symplectic integrator presented in Section~\ref{sectionderivation}
we consider the motion of a particle in an electric field with (scaled) magnetic scalar potential given by:
\begin{eqnarray}
\phi & = & \phi_0 \sqrt{\frac{\cosh(u)-\cos(v)}{\sinh(u)}} 
\left( \frac{1}{12} P_{12-\frac{1}{2}}^{-3}(\coth(u))\sin(12\theta) \right. \nonumber \\
& & \qquad \left. - P_{1-\frac{1}{2}}^{-3}(\coth(u))\sin(\theta) \right) \cos(3v).
\label{testcasemagpotential}
\end{eqnarray}
The field derived from this potential has the characteristics of a skew sextupole field, as shown in
Fig.~\ref{figtestcasemagneticfield}.  We choose the field strength such that $\phi_0 = 5\times 10^4$, and use a
radius of curvature for the reference trajectory $\rho = 5$\,m.  A dipole magnetic field is included, represented
by the longitudinal component of the vector potential (\ref{magneticvectorpotential}), but with $k_0 = 1.05/\rho$
so that there is a slight mismatch between the field and the curvature of the reference trajectory.

For the reference particle, we choose $\beta_0 = 0.8$, and the initial conditions for the particle
to be tracked are:
\begin{eqnarray}
& & (x, p_x, y, p_y, z, \delta) = \nonumber \\
& & \qquad (1\,\textrm{mm}, 4\times 10^{-3}, 1\,\textrm{mm}, -0.1\times 10^{-3}, 0, 0.02). \nonumber \\
& & 
\end{eqnarray}
We track the particle using the explicit symplectic integrator presented in Section~\ref{sectionderivation},
from $s = 0$ to $s = s_\mathrm{max} = \frac{\pi}{6}\rho$, with a step size of $\Delta \sigma = s_\mathrm{max}/10$.
The integration required in (\ref{pxchangegenh1y}) is approximated by Simpson's rule:
\begin{eqnarray}
& & \int_{y_0}^{y_1} \left. \frac{\partial a_y}{\partial x} \right|_{y=\bar{y}} \, d\bar{y} \approx
\frac{y_1 - y_0}{6} \nonumber \\
& & \qquad \times \left(
\left. \frac{\partial a_y}{\partial x} \right|_{y = y_0}
+ 4 \left. \frac{\partial a_y}{\partial x} \right|_{y = \frac{1}{2}(y_0 + y_1)}
+ \left.\frac{\partial a_y}{\partial x} \right|_{y = y_1} \right), \nonumber \\
& & 
\end{eqnarray}
where the derivative is evaluated in each case at the appropriate (fixed) values of $x$ and $s$, and at the
indicated value of $y$.
A similar approximation is made for the integration in (\ref{pychangegenh1x}).
Although these approximations will lead to some symplectic error, this should be small for small step size.  In
cases where symplecticity is important, more accurate integration routines can be used, though at greater
computational cost.

\begin{figure*}
\begin{center}
\includegraphics[width=0.85\textwidth]{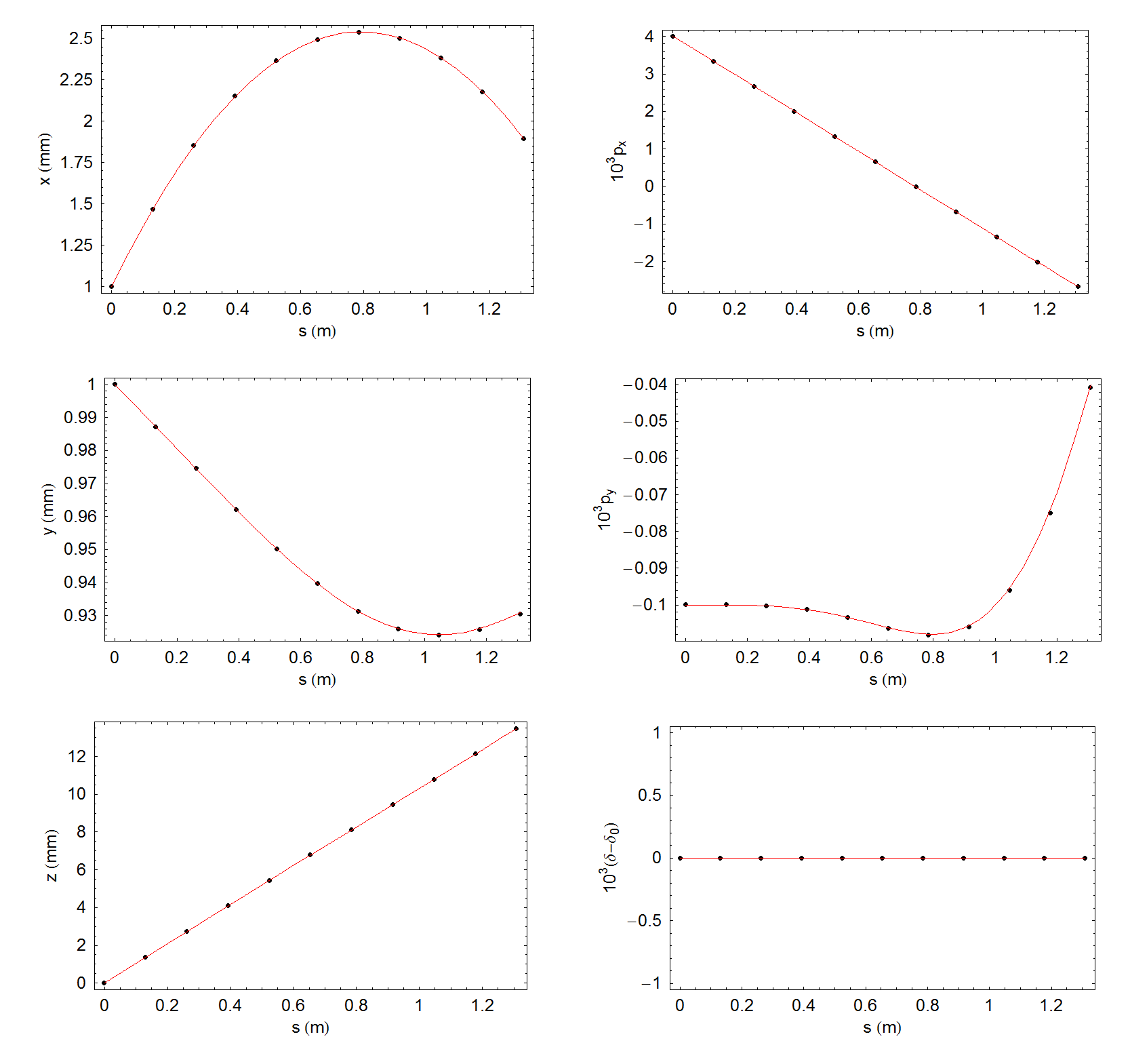}
\caption{Results of tracking a particle through a magnetic curvilinear skew sextupole,
described by the magnetic scalar potential given in Eq.~(\ref{testcasemagpotential}). 
The black points show the results from the explicit symplectic integrator presented in
Section~\ref{sectionderivation}.  The red lines show the results of numerical integration
of the equations of motion derived from the Hamiltonian (\ref{hamiltonian}).\label{figuretestcase0}}
\end{center}
\end{figure*}

The tracking results are shown in Fig.~\ref{figuretestcase0}.  There is good agreement between the
two integration methods.

\subsection{Curvilinear electrostatic quadrupole}
\begin{figure*}
\begin{center}
\includegraphics[width=0.32\textwidth]{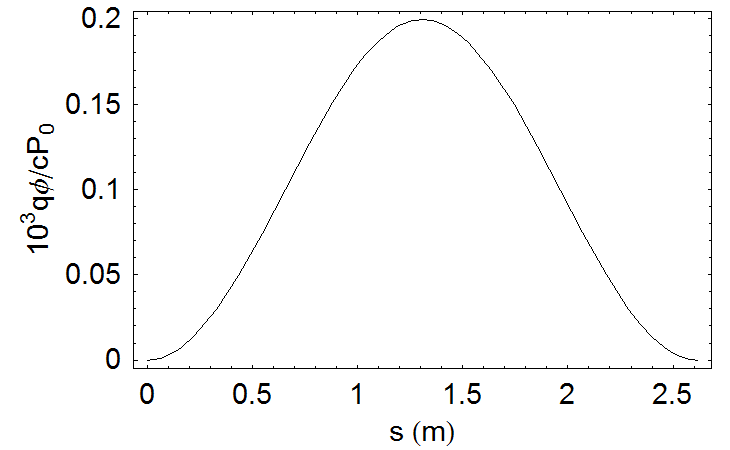}
\includegraphics[width=0.32\textwidth]{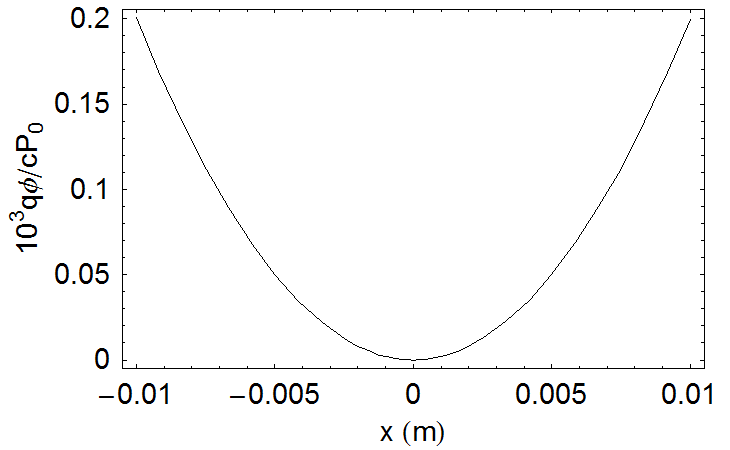}
\includegraphics[width=0.32\textwidth]{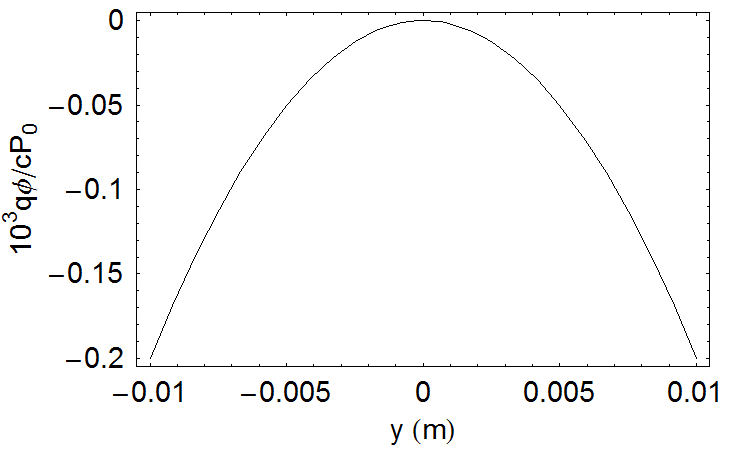}
\caption{Variation of the electrostatic potential (\ref{testcasepotential}) in a curvilinear quadrupole,
as a function of the co-ordinates $s$ (left-hand plot, for $x = 10\,$mm and $y = 0$), $x$ (middle plot,
for $y = 0$ and $s = \frac{\pi}{12}\rho$) and $y$ (right-hand plot, for $x = 0$ and $s = \frac{\pi}{12}\rho$).
\label{figtestcasepotential}}
\end{center}
\end{figure*}

As a second illustration of the explicit symplectic integrator presented in Section~\ref{sectionderivation}
we consider the motion of a particle in an electric field with (scaled) scalar potential given by:
\begin{eqnarray}
\phi & = & \phi_0 \sqrt{\frac{\cosh(u)-\cos(v)}{\sinh(u)}} 
\left( P_{12-\frac{1}{2}}^{-2}(\coth(u))\cos(12\theta) \right. \nonumber \\
& & \qquad \left. - P_{-\frac{1}{2}}^{-2}(\coth(u)) \right) \cos(2v).
\label{testcasepotential}
\end{eqnarray}
This represents the potential for a ``curvilinear'' electrostatic quadrupole, with a strength that varies with
longitudinal position along the reference trajectory.  The transverse and longitudinal variation of the field are
described by $m = 2$ and $n = 12$ (respectively) in Eq.~(\ref{toroidalpotential}).  The potential is
illustrated in Fig.~\ref{figtestcasepotential}.
We choose the field strength $\phi_0 = 200$, and use a radius of curvature for the
reference trajectory $\rho = 5$\,m.  We include a magnetic field, represented by the vector potential
(\ref{magneticvectorpotential}), but we introduce a small mistmatch between the field and the curvature
of the reference trajectory by setting $k_0 = 1.05/\rho$.

\begin{figure*}[ht!]
\begin{center}
\includegraphics[width=0.85\textwidth]{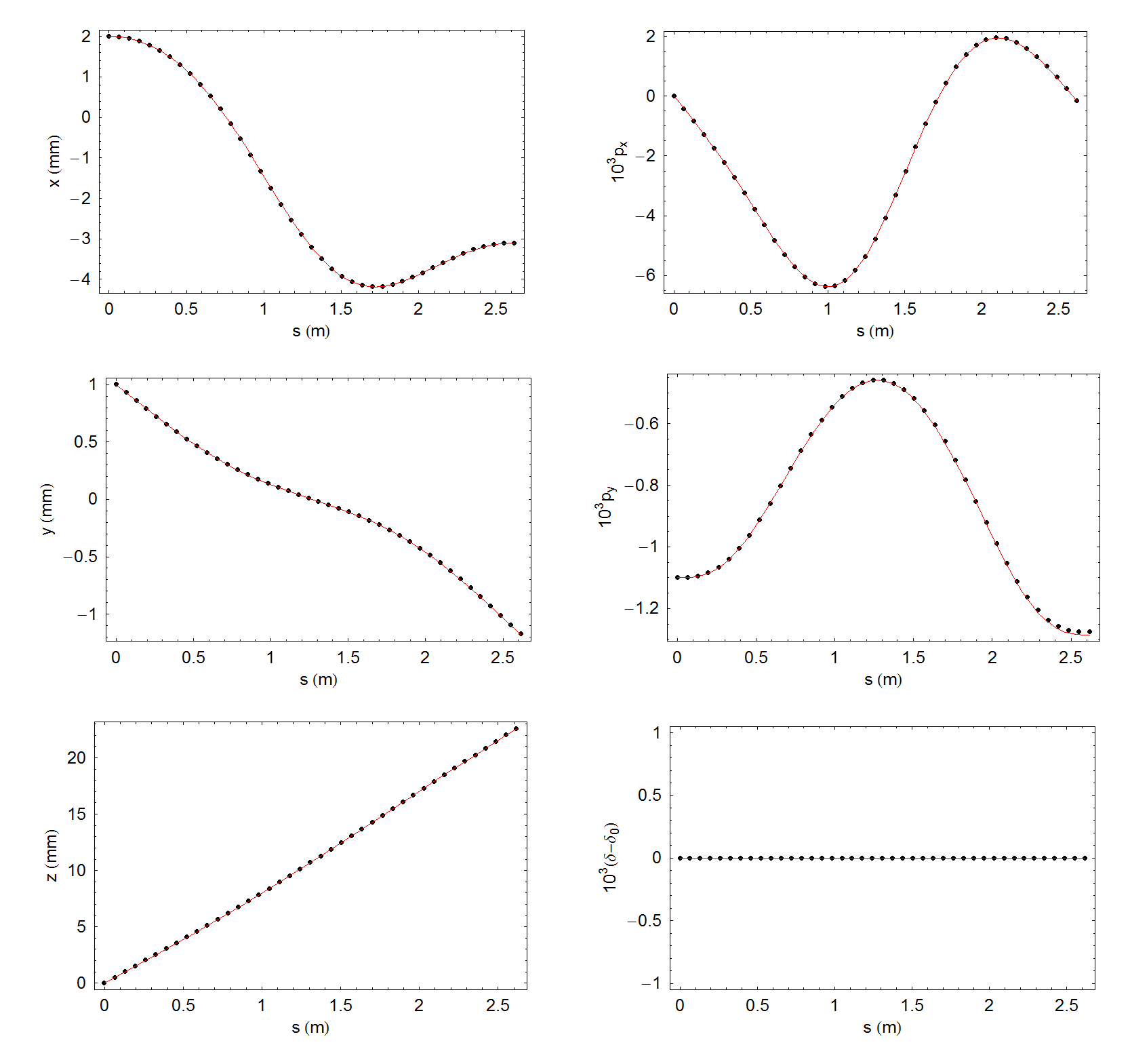}
\caption{Results of tracking a particle through the field of a curvilinear electrostatic quadrupole.
The potential is given by Eq.~(\ref{testcasepotential}).
The black points show the results from the explicit symplectic integrator described in Section~\ref{sectionderivation}.
The red lines show the results from numerical integration of the equations of motion derived
from the Hamiltonian (\ref{hamiltonian}).\label{figuretestcase1}}
\end{center}
\end{figure*}

For the reference particle, we choose $\beta_0 = 0.8$, and the initial conditions for the particle
to be tracked are:
\begin{eqnarray}
& & (x, p_x, y, p_y, z, \delta)= \nonumber \\
& & \qquad (2\,\textrm{mm}, 0, 1\,\textrm{mm}, -1.1\times 10^{-3}, 0, 0.02).
\end{eqnarray}
We track the particle using the explicit symplectic integrator presented in Section~\ref{sectionderivation},
from $s = 0$ to $s = s_\mathrm{max} = \frac{\pi}{6}\rho$, with a step size of $\Delta \sigma = s_\mathrm{max}/40$.
For comparison, we also integrate numerically the (Hamiltonian) equations of motion derived from the exact
Hamiltonian (\ref{hamiltonian}).  The tracking results are shown in Fig.~\ref{figuretestcase1}, and again
we see good agreement between the two integration methods.

\subsection{g-2 storage ring electrostatic quadrupole}
As a final example of application of the symplectic integrator, we consider the fringe field regions of the electrostatic
quadrupoles in the g-2 storage ring \cite{gminus2,gminus2tdrquads,wu2017,semertzidis2003}.
Values for the potential were calculated (using an FEA code) at points on a uniform Cartesian grid; the values of the
potential on a surface defined (in toroidal co-ordinates) by $u = u_\mathrm{ref} = 5.76$ were then obtained by
(spline) interpolation.  On the surface $u = u_\mathrm{ref}$, we used 120 grid points in $v$, with $0 \le v < 2\pi$,
and 80 grid points in $\theta$, with $0 < \theta \le 2^\circ$ (such that the ends of the quadrupole electrodes are
at approximately $\theta = 1^\circ$).  The reference
radius for the co-ordinate system is taken to be the radius of curvature of the reference trajectory in the g-2
storage ring, $\rho = 7.112\,$m: this is the radius of the closed orbit for muons with momentum 3.094\,GeV/c.
The value of $u = 5.76$ then corresponds, for $v = 0$, to a point with $x = 0.045$\,m and $y = 0$, in the
conventional accelerator co-ordinate system, with the origin for the $x$ and $y$ co-ordinates on the reference
trajectory.

Based on equation (\ref{toroidalpotential}), coefficients $f_{mn}$ were calculated so that the potential on any grid
point can be found from:
\begin{eqnarray}
\phi & = & \sqrt{\frac{\cosh(u) - \cos(v)}{\sinh(u)}} \nonumber \\
& & \quad
\times \sum_{m, n} f_{mn} i^m P_{n^\prime - \frac{1}{2}}^{-m}(\coth(u)) \cos(mv) \sin(n^\prime \theta),
\nonumber \\
& & 
\label{gm2quadrupolepotentialfit}
\end{eqnarray}
where $n^\prime = n_0(2n+1)$, with $n_0 = 45$ (so that $n = 0$ corresponds to a sine function with quarter
period equal to $2^\circ$, i.e.~the range of $\theta$ over which values for the potential are given).
The values of $f_{mn}$ are obtained essentially by a discrete Fourier transform of the potential on
the given grid points.  Mode numbers $0 \le m \le 10$ and $0 \le n \le 79$ are used.  The truncation in
the azimuthal mode number $m$ (compared to the number of data points available) means that the data
are not fitted perfectly; however, the contribution of modes (multipoles) of order $m > 10$ is found to be small.
Note that the dominant multipole is the quadrupole component, $m = 2$.

\begin{figure}
\begin{center}
\includegraphics[width=\columnwidth]{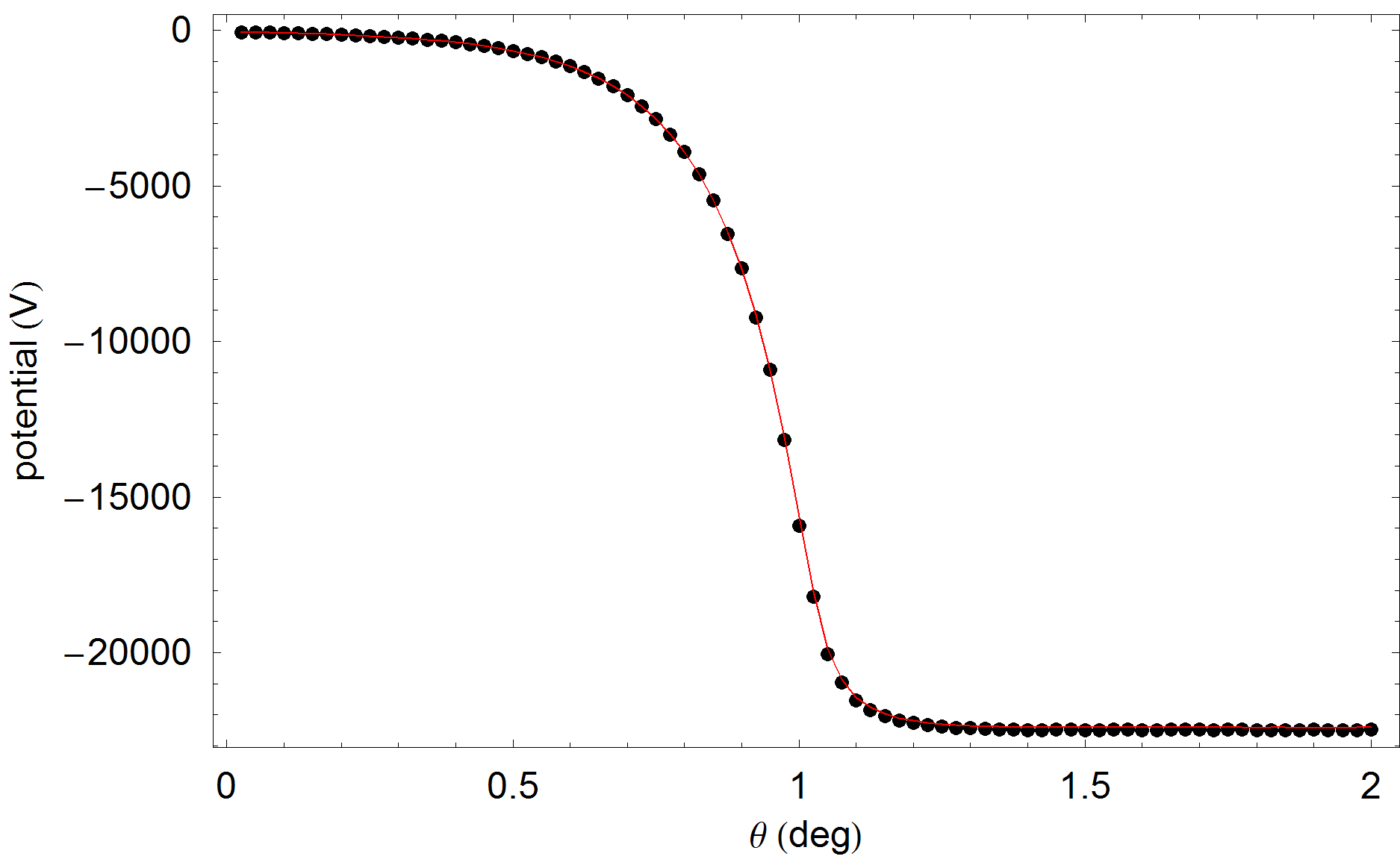}
\caption{Scalar potential in an electrostatic quadrupole in the g-2 storage ring.  The potential is plotted as a function
of toroidal co-ordinate $\theta$ at $u = u_\mathrm{ref}$ and $v = 0$.  The black points show the original data points; the
red line shows a fit using equation (\ref{gm2quadrupolepotentialfit}).\label{figuregm2potentialfit2}}
\end{center}
\end{figure}

\begin{figure*}
\begin{center}
\includegraphics[width=0.48\textwidth]{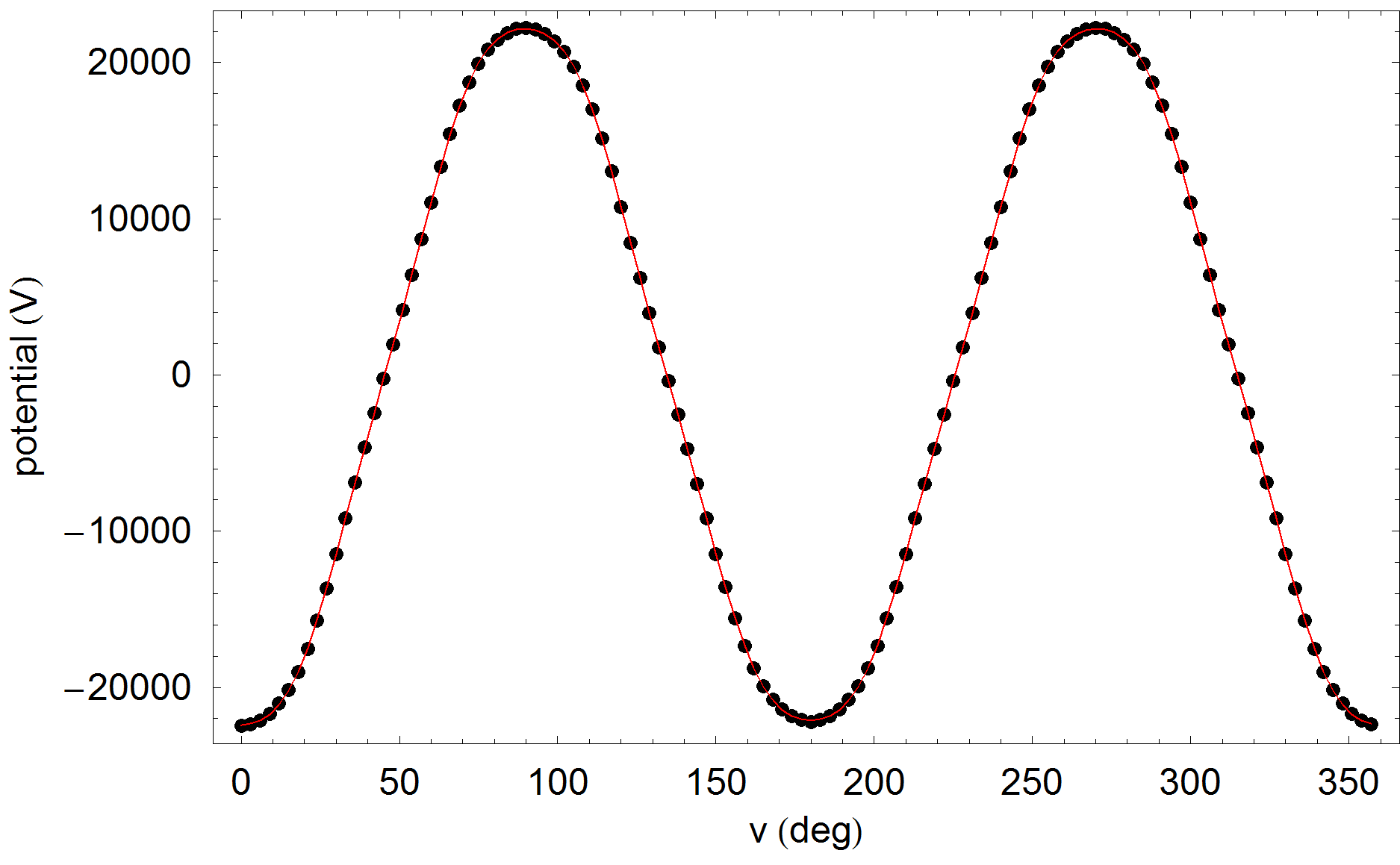}
\includegraphics[width=0.48\textwidth]{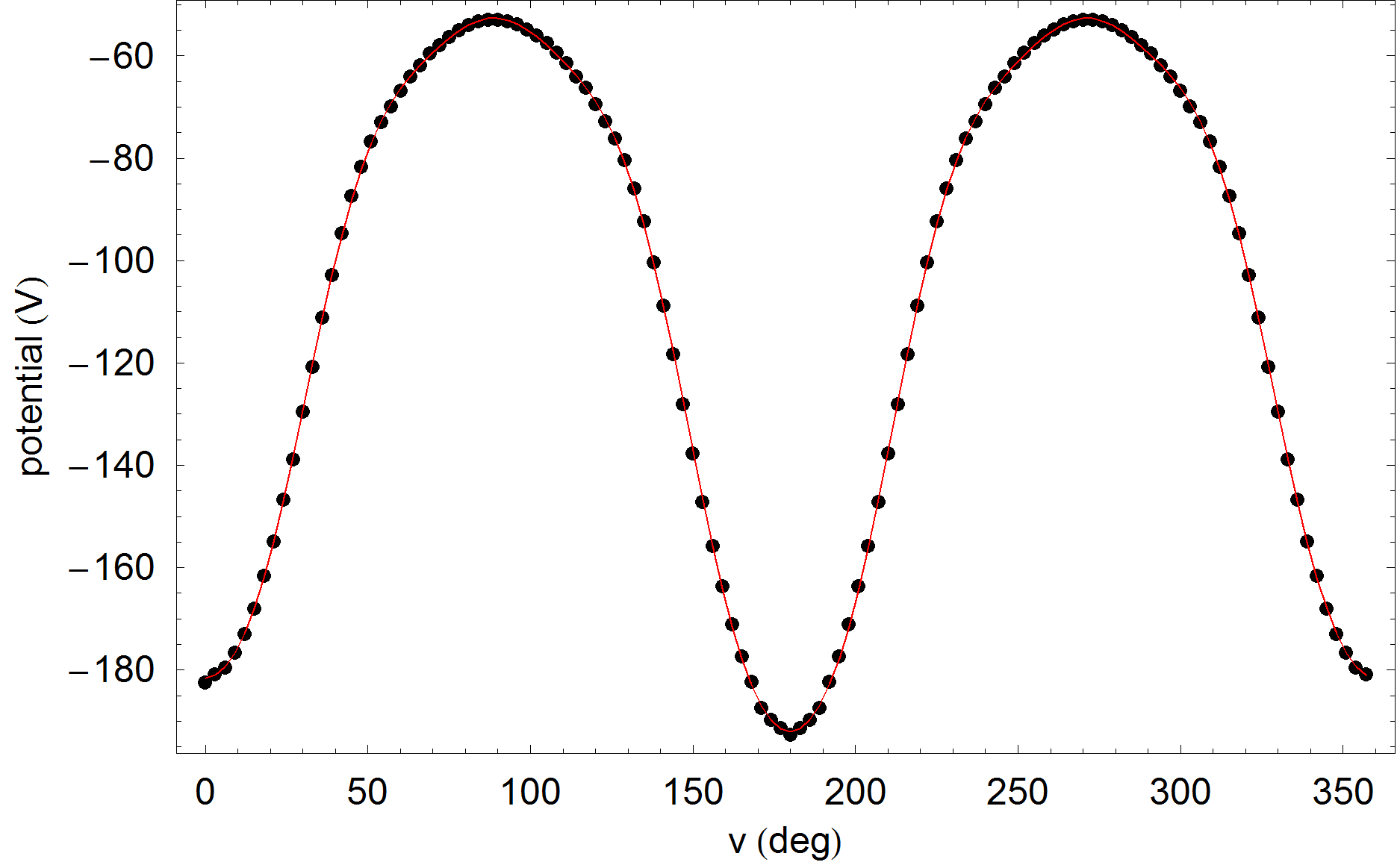}
\caption{Scalar potential in an electrostatic quadrupole in the g-2 storage ring.  The potential is plotted as a function
of toroidal co-ordinate $v$ at $\theta = 2^\circ$ (left) and at $\theta = 0.25^\circ$ (right), with
$u = u_\mathrm{ref}$ in both cases.  The black points show the original data points; the
red lines show fits using equation (\ref{gm2quadrupolepotentialfit}).\label{figuregm2potentialfit1}}
\end{center}
\end{figure*}

The potential as a function of $\theta$ (at $u = u_\mathrm{ref}$ and $v = 0$) is shown in Fig.~\ref{figuregm2potentialfit2},
and as a function of $v$ (at $\theta = 2^\circ$ and at $\theta = 0.25^\circ$, with $u = u_\mathrm{ref}$ in
both cases) in Fig.~\ref{figuregm2potentialfit1}.  In the lower plot in Fig.~\ref{figuregm2potentialfit1}, we see that
the variation of the potential with the ``azimuthal'' co-ordinate $v$ in the fringe-field region (about 30\,mm from the ends
of the electrodes) is significantly distorted from a simple sine wave, indicating the presence of higher-order multipoles.

\begin{figure*}
\begin{center}
\includegraphics[width=0.48\textwidth]{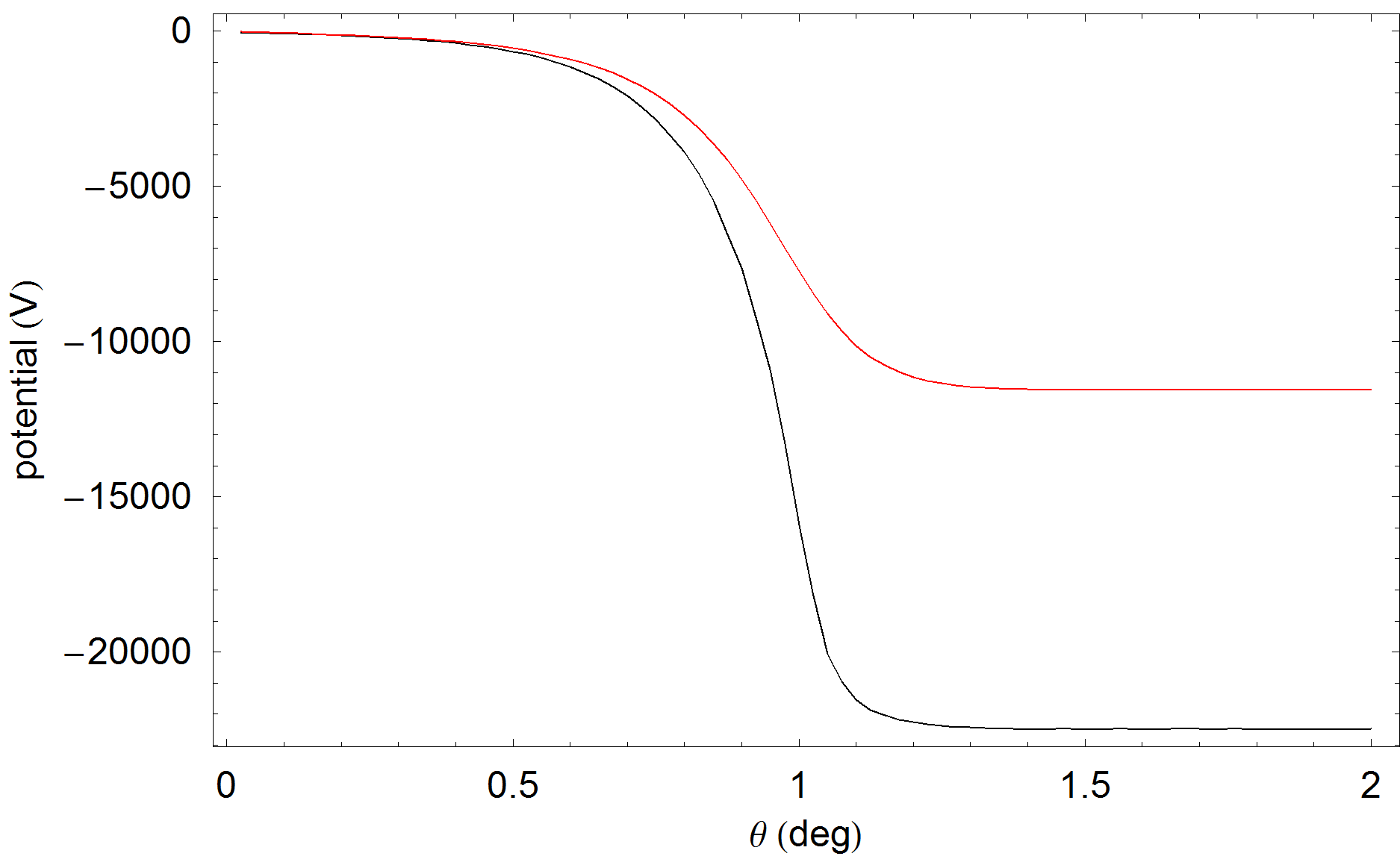}
\includegraphics[width=0.48\textwidth]{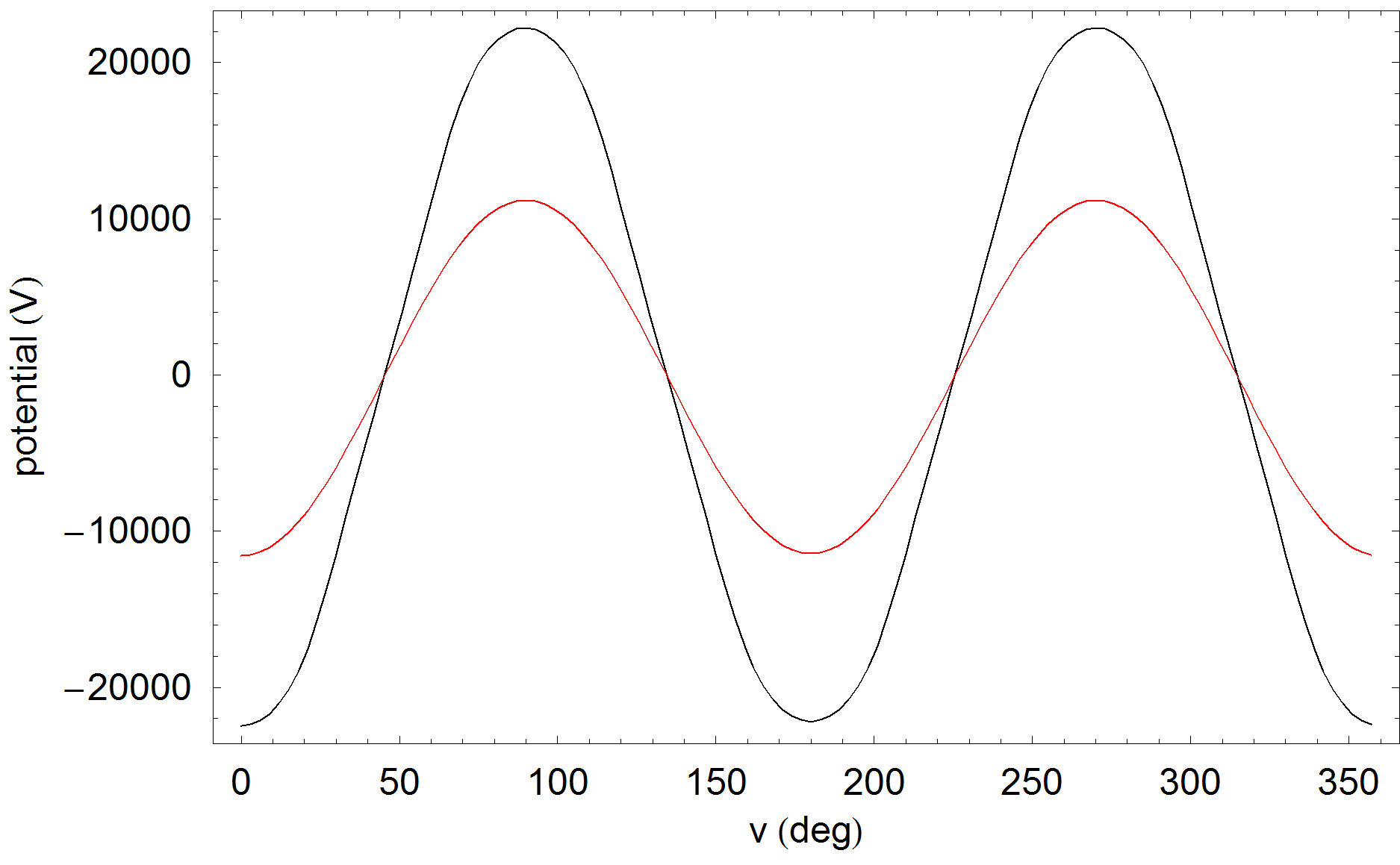}
\caption{Scalar potential in an electrostatic quadrupole in the g-2 storage ring.  The potential is plotted as a function
of toroidal co-ordinate $\theta$ at $v = 0$ (left) and as a function of $v$ at $\theta = 2^\circ$ (right).
In each plot, the black line shows the potential at $u = u_\mathrm{ref} = 5.76$,
and the red line shows the potential at $u = 6.11$.  At the larger value of $u$, the value of the co-ordinate $x$ is
reduced by a factor of $\sqrt{2}$ compared to the value of $x$ at $u = u_\mathrm{ref}$; the potential is a factor
of two smaller at the larger value of $u$, as expected for a quadrupole field.\label{figuregm2potentialfit4}}
\end{center}
\end{figure*}

Using the coefficients $f_{mn}$ we can calculate the potential at any point within the surface on which the
fit is performed.  As an example, Fig.~\ref{figuregm2potentialfit4} shows the potential as a function of
$\theta$ (for $v = 0$) and as a function of $v$ (for $\theta = 2^\circ$).  In each plot, the black line shows
the potential at $u = u_\mathrm{ref} = 5.76$ and the red line shows the potential at $u = 6.11$: the larger
value of $u$ corresponds to a value of $x$ that is a factor of $\sqrt{2}$ smaller than the value of $x$ at
$u = u_\mathrm{ref}$, so that the potential (for a pure quadrupole) is expected to be smaller by a factor of
two.  The expected behaviour of the potential (as a function of $u$) is indeed what we observe.

\begin{figure*}
\begin{center}
\includegraphics[width=0.85\textwidth]{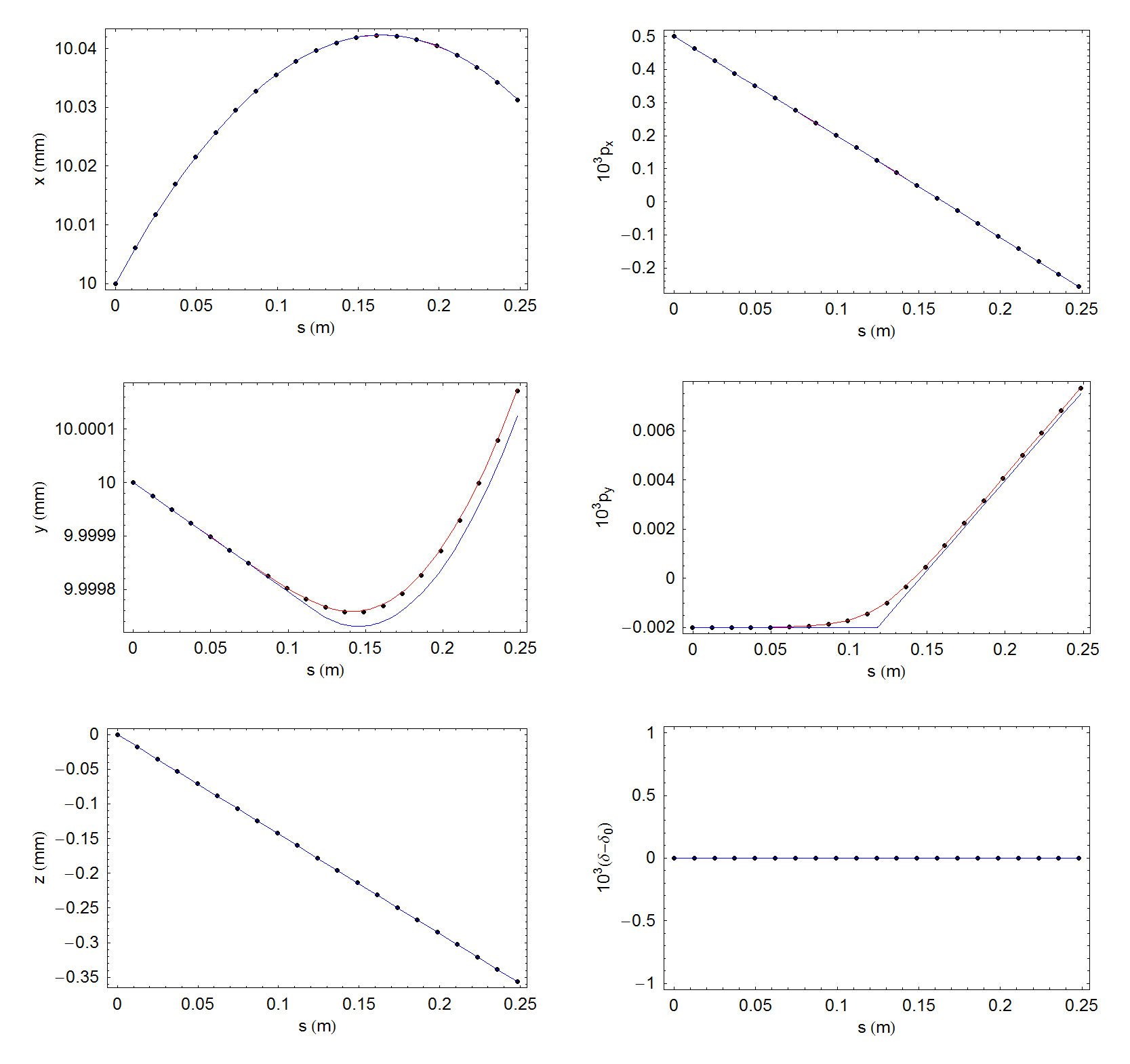}
\caption{Trajectory of a muon through the fringe field region of an electrostatic quadrupole in the g-2 storage
ring.  The electrostatic potential is shown in Figs.~\ref{figuregm2potentialfit2} and \ref{figuregm2potentialfit1}.
The reference momentum is 3.094\,GeV/c, and the reference trajectory is the arc of a circle with radius 7.112\,m,
determined by the magnetic field strength, $B \approx 1.45\,$T.  The initial co-ordinates $(x,p_x,y_py,z,\delta)$ of
the muon are $(10\,\textrm{mm},5\times 10^{-4}, 10\,\textrm{mm}, -2\times 10^{-6}, 0 , -0.02)$.
The black points show the results from the symplectic integrator, with step size 12.4\,mm, i.e.~a total of 20
steps.  The red line shows the results of an integration using a (non-symplectic) adaptive Runge--Kutta integration
of the equations of motion in the same field.  The blue line shows the results
of a Runge--Kutta integration of the equations of motion through a region with the same magnetic field, but with
a ``hard-edge'' model for the electric field. \label{figuregm2tracking}}
\end{center}
\end{figure*}

Tracking a particle through the fringe field of an electrostatic quadrupole using the symplectic integrator described
in Section \ref{sectionderivation} requires the derivatives of the potential with respect to the accelerator co-ordinates,
$x$, $y$ and $s$.  The derivatives can be calculated (at any point within the surface used to fit the coefficients
$f_{mn}$ for the given potential) using equation (\ref{gm2quadrupolepotentialfit}), together with (\ref{dudx})
and (\ref{dudy}).  Some example results from tracking a muon through the fringe field are shown in
Fig.~\ref{figuregm2tracking}.  The black points in Fig.~\ref{figuregm2tracking} show the muon trajectory calculated
using the symplectic integrator for the detailed fringe-field model, i.e.~the model based on the numerical data for the
scalar potential.  The red line shows the results of an integration using a (non-symplectic) adaptive Runge--Kutta
integration of the equations of motion in the same field.  The blue line shows the results of a Runge--Kutta integration
of the equations of motion through a region with the same magnetic field, but with a ``hard-edge'' model for the
electric field.  The hard-edge model is constructed so that the scalar potential is zero up to a point $s = s_1$, and is
given simply by $\phi = \frac{1}{2}k_1(x^2 - y^2)$ for $s > s_1$.  The value of $k_1$ is chosen to correspond to
the focusing potential in the body of the quadrupole found from the numerical data for the scalar potential.  The
point $s_1$ is chosen so that the integrated gradient, $\int_0^{s_\mathrm{max}} k_1\,ds$ in the
hard-edge model is equal to the integrated gradient in the fringe-field model.

There is good agreement between the symplectic integrator and the Runge--Kutta integrator for the detailed fringe-field
model.  There is little difference between the detailed fringe-field model and the hard-edge model for the horizontal motion,
which is dominated by the magnetic field (that is the same in both cases).  There is some small but observable
difference between the detailed fringe-field model and the hard-edge model for the vertical motion.  The change in the vertical
momentum after integrating through the full region is approximately the same in both cases: this is expected, since
the length of the quadrupole field in the hard-edge model was chosen to give the same integrated focusing strength
as the detailed fringe-field model.  However, the fact that the change in the vertical momentum occurs at a discrete
point in the hard-edge model leads to a slightly larger difference between the models in the vertical co-ordinate at the
end of the integration.  It is unclear what impact this may have on the beam dynamics in the storage ring, but it is
possible that it may lead to an observable effect over a sufficiently large number of turns.
\vspace*{0.5cm}

\begin{acknowledgments}
We wish to thank Bruno Muratori for useful discussions, and members of the g-2 Collaboration for help and support with
studies of the g-2 electrostatic quadrupoles.  In particular, we are grateful to Wanwei Wu for providing field data.
\end{acknowledgments}

\end{document}